\documentclass[preprint,showpacs,preprintnumbers,amsmath,amssymb]{revtex4}

\usepackage{graphicx}
\usepackage{dcolumn}
\usepackage{bm}

\begin{document}

\preprint{APS/123-QED}

\title{A Modified Scalar-Tensor-Vector Gravity Theory and the Constraint on its Parameters}

\author{Xue-Mei Deng$^{1}$}
 \email{xue.mei.deng.x@gmail.com}
\author{Yi Xie$^{2}$}
 \email{yixie@nju.edu.cn}
\author{Tian-Yi Huang$^{2,3}$}
 \email{tyhuang@nju.edu.cn}
 \affiliation{%
$^{1}$Purple Mountain Observatory, Chinese Academy of Sciences,
Nanjing 210008, China\\
$^{2}$Department of Astronomy, Nanjing University, Nanjing 210093,
China\\
$^{3}$Shanghai Astronomical Observatory, Chinese Academy of
Sciences, Shanghai 200030, China }%

\date{\today}

\begin{abstract}
A gravity theory called scalar-tensor-vector gravity (STVG) has been
recently developed and succeeded in solar system, astrophysical and
cosmological scales without dark matter [J. W. Moffat, J. Cosmol.
Astropart. Phys. 03, 004 (2006)]. However, two assumptions have been
used: (i) $B(r)=A^{-1}(r)$, where $B(r)$ and $A(r)$ are $g_{00}$ and
$g_{rr}$ in the Schwarzschild coordinates (static and spherically
symmetric); (ii) scalar field $G=Const.$ in the solar system. These
two assumptions actually imply that the standard parametrized
post-Newtonian parameter $\gamma=1$. In this paper, we relax these
two assumptions and study STVG further by using the post-Newtonian
(PN) approximation approach. With abandoning the assumptions, we
find $\gamma\neq1$ in general cases of STVG. Then, a version of
modified STVG (MSTVG) is proposed through introducing a coupling
function of scalar field G: $\theta(G)$. We have derived the metric
and equations of motion (EOM) in 1PN for general matter without
specific equation of state and $N$ point masses firstly.
Subsequently, the secular periastron precession $\dot{\omega}$ of
binary pulsars in harmonic coordinates is given. After discussing
two PPN parameters ($\gamma$ and $\beta$) and two Yukawa parameters
($\alpha$ and $\lambda$), we use $\dot{\omega}$ of four binary
pulsars data (PSR B1913+16, PSR B1534+12, PSR J0737-3039 and PSR
B2127+11C) to constrain the Yukawa parameters for MSTVG:
$\lambda=(3.97\pm0.01)\times10^{8}$m and
$\alpha=(2.40\pm0.02)\times10^{-8}$ if we fix $|2\gamma-\beta-1|=0$.
\end{abstract}

\maketitle

\section{Introduction}

With tremendous advance in the accuracy of observations, Einstein's
general relativity (GR) has passed nearly all the tests in the solar
system. However, alternative gravity theories still stand up for
explaining some exotic phenomena such as the dark matter and for
possible violation of general relativity in the future higher
precision experiments. Among them, Moffat \cite{b7} proposed a
scalar-tensor-vector gravity (STVG) theory in which there are three
scalar fields and a vector field besides the metric tensor. Three
scalar fields are respectively the scalar field $G$ that origins
from the Newtonian gravitational constant, the coupling function of
the vector field $\omega$ and the rest mass of the vector field $m$
which controls the coupling range. They all change with space and
time. The vector field $\phi_{\mu}$, which is associated with a
fifth force charge, corresponds to the exchange of a massive spin 1
boson and couples to the ordinary matter. Through introduced
$\phi_{\mu}$, a Yukawa-like force was added to the Newtonian inverse
square law. This leads to a satisfied fit to galaxy rotation curves
and the Tully-Fisher law \cite{b8}. Besides, the theory has been
used successfully to explain cosmological observations \cite{b9},
the motion of galaxy clusters \cite{b10}, the Bullet Cluster
\cite{b11}, the velocity dispersion profiles of satellite galaxies
\cite{b12} and globular clusters \cite{b13} without exotic dark
matter. By studying STVG attentively, we found it uses two
assumptions in STVG:
\begin{enumerate}
  \item The metric components $g_{00}$ and $g_{rr}$ in a spherically symmetric field have
the following relationship: $B(r)=A^{-1}(r)$ , where $B(r)$ and
$A(r)$ are $g_{00}$ and $g_{rr}$ respectively in a
Schwarzschild-like coordinate system;
  \item For the solar system, the running of gravitational constant $G$
is zero, namely, $G=Const$.
\end{enumerate}
With the above assumptions, let us analysis what results they lead
to. A general form of the metric in the Schwarzschild coordinates
reads
\begin{equation}
\label{metric1} \mathrm{d}s^{2}=-B(r)c^{2}\mathrm{d}t^{2}
+A(r)\mathrm{d}r^{2}+r^{2}\mathrm{d}^{2}\Omega,
\end{equation}
where
$\mathrm{d}^{2}\Omega=\mathrm{d}\theta^{2}+\sin^{2}\theta\mathrm{d}\varphi^{2}$.
Comparing Eq. (\ref{metric1}) with the following parametrized
post-Newtonian metric,
\begin{equation}
\label{metric}
\mathrm{d}s^{2}=-\bigg(1-2\frac{GM}{c^{2}r}\bigg)c^{2}\mathrm{d}t^{2}
+\bigg(1+2\gamma\frac{GM}{c^{2}r}\bigg)\mathrm{d}r^{2}+r^{2}\mathrm{d}^{2}\Omega,
\end{equation}
within the post-Newtonian precision $\mathcal{O}(1/c^{2})$, it is
clear that $\gamma=1$ under assumption (i). On the other hand, since
only the scalar field $G$ affects $\gamma$ in STVG (for details see
subsection D of section II), assumption (ii) gives $\gamma=1$
directly. In summary, these assumption impose a constraint on the
theory: the standard parametrized post-Newtonian (PPN) parameter
$\gamma$ equals to 1. Therefore, these two assumptions are
reasonable and consistent with the current measurement of $\gamma$.

However, we prefer to a general form in natural science without any
imposed limitations. For the first assumption, we have no strong
reason to take it. For the second assumption, the search for the
Newtonian gravitational constant never stops. For instance,
planetary and spacecraft ranging, neutron star binary observations,
paleontological and primordial nucleosynthesis data allow one to
constrain the variation of $G$ with time \cite{100}. This is the
reason why we relax these two assumptions and study STVG further.
Firstly, Let us investigate the numerical value of $\gamma$ in STVG
attentively with abandoning these two assumptions in comparison with
Brans-Dicke theory. We expand metric as follows
\begin{eqnarray}
\label{g00}
g_{00}&=&-1+\epsilon^{2}N+\epsilon^{4}L+\mathcal {O}(5),\\
\label{g0i}
g_{0i}&=&\epsilon^{3}L_{i}+\mathcal {O}(5),\\
\label{gij} g_{ij}&=&\delta_{ij}+\epsilon^{2}H_{ij}+\mathcal {O}(4),
\end{eqnarray}
by using Chandrasekhar's approach \cite{b14}, we have found $N$ and
$H_{ij}$ only depend on the scalar field $G$ and matter in STVG (for
details see subsection D of section II) and where $\epsilon=1/c$ and
$\mathcal{O}(n)$ means of order $\epsilon^{n}$. Then, we can compare
STVG with Brans-Dicke theory (BD) \cite{b22}. The action of BD is
\begin{equation}
\label{BD} \mathcal {S}_{BD}=\frac{c^{3}}{16\pi}\int\bigg(\phi
R-\varsigma_{0}g^{\mu\nu}\frac{\phi_{;\mu}\phi_{;\nu}}{\phi}\bigg)\sqrt{-g}d^{4}x+\mathcal
{S}_{M},
\end{equation}
where $\varsigma_{0}$  is a coupling constant ($\omega_{0}$ usually
is used in BD, here we use $\varsigma_{0}$ to avoid confusion with
another symbol in the context). For STVG \cite{b7,b9,77,SET}, the
action by only considering the metric and $G$ is
\begin{equation}
\label{stvg} \mathcal
{S}_{STVG}=\frac{c^{3}}{16\pi}\int\bigg(\frac{1}{G}
R+\frac{1}{2}g^{\mu\nu}\frac{G_{;\mu}G_{;\nu}}{G^{3}}\bigg)\sqrt{-g}d^{4}x+\mathcal
{S}_{M}.
\end{equation}
Comparing Eq. (\ref{BD}) with Eq. (\ref{stvg}), we have $G=1/\phi$
and $\varsigma_{0}=-1/2$. In BD,
$\gamma\equiv(\varsigma_{0}+1)/(\varsigma_{0}+2)$ and when
$\varsigma_{0}\rightarrow\infty$, $\gamma=1$. With the
correspondance between BD and STVG, we obtain
$\gamma\equiv(\varsigma_{0}+1)/(\varsigma_{0}+2)=1/3\neq1$ in
Moffat's STVG. (But in some Refs.\cite{45,46,47}, the value of
$\varsigma_{0}$ is $1/2$ due to the sign of the scalar field action
changed and correspondent $\gamma$ is $3/5$.) In subsection D of
section II and Appendix A, we give a strict proof about
$\gamma\neq1$ of STVG in general cases.

Although STVG is a theory that violates EEP, with relaxing the above
two assumptions, the departure of $\gamma$ from 1 should not be out
of the range restricted by current experiments. For example, the
measurement of $\gamma$ in the Cassini experiment gives
$\gamma-1=(2.1\pm2.3)\times10^{-5}$ \cite{b41}. The data analysis
for this result is conducted under standard PPN framework. One might
argue that STVG modifies the Newtonian law by introducing a Yukawa
potential, which breaks the PPN framework and therefore the
constraint given by Cassini may not be used directly here. Another
measurement of $\gamma$ comes from the Lense-Thirring effect
(precession of a gyroscope located near a rotating massive body).
Lense-Thirring effect will cause advance of the ascending node of
the orbit of an earth satellite, which depends only on a force
perpendicular to its orbital plane. But the Yukawa force is confined
in the orbital plane so that the measurements of Lense-Thirring
effect can give a clear and direct constraint on $\gamma$ in STVG.
Although the Lense-Thirring effect for LAGEOS and LAGEOSII are at a
precision level of $10\%$ \cite{b23}, worse than the Cassini
experiment, this experiment demands that $\gamma$ can not departs
from 1 too much (We will explain this in subsection A of section III
in detail). This is the main motivation to propose a modified STVG
theory in this paper when we abandon these two assumptions. It is
more competitive for present and future experiments.

With discovery of the binary pulsar PSR B1913+16 by Hulse and Taylor
\cite{b24}, binary pulsars promises an unprecedented opportunity to
measure the effects of relativistic gravitation (see \cite{70,b45}
for a review). For example, pulsar timing has provided indirect
evidence for the existence of gravitational waves \cite{b16}, the
binary pulsars data can constrain the existence of massive black
hole binaries \cite{b17}, and the binary pulsars can also test the
effects of strong relativistic internal gravitational fields on
orbital dynamics \cite{b18}. In addition, binary pulsars could help
us to test various gravity theories. By fitting the arrival time of
pulsars, observational parameters of binary pulsar are obtained in
high precision. They are ``physical" parameters, ``Keplerian"
parameters and ``post-Keplerian" parameters \cite{b26,b15}. A very
important class is ``post-Keplerian" parameters \cite{b60,b61}
which include the average rate of periastron shift $\dot{\omega}$,
redshift dilation $\gamma$, orbital period derivative $\dot{P}_{b}$,
and two Shapiro-delay parameters $s$ and $r$. It worthy of noted
that the periastron advance for binary pulsars could reach several
degrees per year, which is about $10^{5}$ more than the perihelion
advance of Mercury. Hence, the relativistic effects from binary
pulsar are more remarkable than other celestial systems. We will
take a sample of binary pulsars for testing MSTVG by using their
$\dot{\omega}$. In this paper we chose four best studied pulsar
binaries. They are respectively PSR B1913+16, PSR B1534+12, PSR
J0737-3039, and PSR B2127+11C, which are double neutron star
binaries. We mainly focus on $\dot{\omega}$ of these four binaries
data to constrain the parameters of MSTVG.

Through introducing a coupling function of the scalar field $G$:
$\theta(G)$, we obtain the metric and equations of motion (EOM) for
general matter without specific equation of state and $N$ point
masses in the first order post-Newtonian (1PN) approximation by
Chandrasekhar's approach. Then, the secular periastron precession of
binary pulsars for 1PN in harmonic coordinates is derived. After two
PPN parameters ($\gamma$ and $\beta$) and two Yukawa parameters
($\alpha$ and $\lambda$) are discussed and compared, we fix
$\gamma=1$ and $\beta=1$ and constrain the Yukawa parameters. This
constraint coming from binary pulsars systems on $\alpha$ and
$\lambda$ are consistent with the results from the solar system such
as the earth-satelite measurement of earth gravity, the lunar
orbiter measurement of lunar gravity, and lunar laser ranging
measurement to constrain the fifth force.

In what follows, our conventions and notations generally follow
those of \cite{33}, the metric signature is (-, +, +, +). Greek
indices take the values from 0 to 3, while Latin indices take the
values from 1 to 3. A comma denotes a partial derivative, and
semicolon denotes a covariant derivative. Bold letters denote
spatial vectors. The plot of this paper is as follows. In the next
section, the equations of motion and $\dot{\omega}$ for binary of
MSTVG in 1PN are given. Subsequently, in the third section, we
discuss parameters of MSTVG in detail. We then constrain the
parameters of MSTVG by means of fitting $\dot{\omega}$ for four
binary pulsars and deal with the details of discussion about results
in the forth section. Finally, constraints method and results are
outlined in the last section.

\section[]{The theory}
\subsection[]{Action and field equations}
Through introducing a coupling function of the scalar field $G$:
$\theta(G)$, we adopt the following modified action for STVG
\begin{equation}
\label{action} \mathcal {S}=\int(\mathcal {L}_{G}+\mathcal
{L}_{\phi}+\mathcal {L}_{s})\sqrt{-g}d^{4}x+\mathcal
{S}_{M}(g_{\mu\nu},\phi_{\mu},\Psi),
\end{equation}
where $\Psi$ denotes all the matter fields. In Eq. (\ref{action}),
the matter fields $\Psi$ interact with both the metric field and the
vector field. This means that the trajectory of a free-fall test
particle depends not only on the space-time geometry but also on the
vector field so that it violates the Einstein equivalence principle
(EEP). Although the current experiments verify EEP to a very high
accuracy in the Solar System scale, violations of EEP at galactic
and cosmological distance scales can not be ruled out. In Eq.
(\ref{action}), the Lagrangian densities of the gravitational field,
vector field and scalar fields are respectively
\begin{eqnarray}
\label{tensor} \mathcal{L}_{G}&=&\frac{c^{3}}{16\pi G}(R+2\Lambda),\\
\label{vector1} \mathcal
{L}_{\phi}&=&-\frac{1}{c}\omega\bigg[\frac{1}{4}B^{\mu\nu}B_{\mu\nu}
+\frac{1}{2}m^{2}k^{2}\phi_{\mu}\phi^{\mu}-V_{\phi}(\phi)\bigg],\\
\label{scalar} \mathcal {L}_{s}&=&-\frac{c^{3}}{16\pi
G}\bigg[\frac{1}{2}g^{\mu\nu}\bigg(\theta(G)\frac{G_{;\mu}G_{;\nu}}{G^{2}}
-\frac{m_{;\mu}m_{;\nu}}{m^{2}}+\omega_{;\mu}\omega_{;\nu}\bigg)
-\frac{V_{G}(G)}{G^{2}}+\frac{V_{m}(m)}{m^{2}}-V_{\omega}(\omega)\bigg],
\end{eqnarray}
where $\Lambda$ denotes cosmological constant and $V_{X}(X)$ denotes
the self-interaction potential function of an field. Besides,
$k=c^{2}/\hbar^{2}$ and $B_{\mu\nu}=\phi_{\nu,\mu}-\phi_{\mu,\nu}$.
Where $c$ is the ultimate speed of the special theory of relativity
and $\hbar$ is the reduced Planck constant. It is noted that in
Moffat's STVG \cite{b7,b9,77,SET}, $\theta(G)=-1$ in Eq.
(\ref{scalar}).

When we omit $\Lambda$ and all the self-interaction potentials
$V_{X}(X)$, equations of gravitational field are obtained by
variation of the action (\ref{action}) with respect to $g^{\mu\nu}$.
\begin{eqnarray}
\label{R} R_{\mu\nu}=&&\frac{8\pi
G}{c^{2}}\bigg[T_{\mu\nu}-\frac{1}{2}g_{\mu\nu}T+\omega\frac{m^{2}k^{2}}{c^{2}}\phi_{\mu}\phi_{\nu}
+\frac{1}{c^{2}}\omega\bigg(B_{\mu\kappa}B_{\nu}^{~\kappa}-\frac{1}{4}g_{\mu\nu}B_{\kappa\lambda}B^{\kappa\lambda}\bigg)\bigg]\nonumber\\
&&+\frac{1}{2}\bigg(\theta(G)\frac{G_{;\mu}G_{;\nu}}{G^{2}}+4\frac{G_{;\mu}G_{;\nu}}{G^{2}}-\frac{m_{;\mu}m_{;\nu}}{m^{2}}
-2\frac{G_{;\mu\nu}}{G}+\omega_{;\mu}\omega_{;\nu}\bigg)\nonumber\\&&
-\frac{1}{2}g_{\mu\nu}\bigg(\frac{G_{;\kappa}^{~;\kappa}}{G}-2\frac{G_{;\kappa}G^{;\kappa}}{G^{2}}\bigg),
\end{eqnarray}
where $T_{\mu\nu}$ is the stress-energy-momentum tensor of matter
which is defined by
\begin{equation}
\frac{c}{2}\sqrt{-g}T_{\mu\nu}\equiv-\frac{\delta\mathcal
{S}_{M}}{\delta g^{\mu\nu}}.
\end{equation}
Following \cite{b46} and \cite{b47,b4}, we define mass, current,
and stress density as
\begin{eqnarray}
\label{sigma}
\sigma&\equiv&T^{00}+T^{kk},\\
\label{sigmai}
\sigma_{i}&\equiv&cT^{0i},\\
\label{sigmaij} \sigma_{ij}&\equiv&c^{2}T^{ij}.
\end{eqnarray}
It is worth emphasizing that, in these definitions, the matter is
described by the energy-momentum tensor without specific equation of
state.

Variation of the action with respect to $\phi^{\mu}$ yields
\begin{equation}
\label{vector2} \omega
B^{\nu\mu}_{;\nu}+B^{\nu\mu}\omega_{;\nu}-\omega
m^{2}k^{2}\phi^{\mu}=\mathcal{J}^{\mu},
\end{equation}
where $\mathcal{J}^{\mu}$ is a ``fifth force" matter current defined
as
\begin{equation}
\label{eep}
\sqrt{-g}\frac{\mathcal{J}^{\nu}}{c}\equiv-\frac{\delta\mathcal
{S}_{M}}{\delta \phi_{\nu}}.
\end{equation}
We further define
\begin{equation}
\label{J1} \mathcal{J}^{0}\equiv J^{0},~~~~\mathcal{J}^{i}\equiv
\epsilon J^{i}.
\end{equation}
When we substitute $k=c^{2}/\hbar^{2}$ and
$B_{\mu\nu}=\phi_{\nu,\mu}-\phi_{\mu,\nu}$ into Eq. (\ref{vector2}),
we can obtain the Proca equation if $\omega=Const.$ and
$\mathcal{J}^{\mu}=0$.

Variation with respect to the scalar fields yield respectively
\begin{eqnarray}
\label{G}
&&(\theta(G)+3)\frac{1}{G}G^{;\kappa}_{~;\kappa}-2(\theta(G)+3)\frac{G_{;\kappa}G^{;\kappa}}{G^{2}}
-\frac{1}{2}\frac{\mathrm{d}\theta(G)}{\mathrm{d}G}\frac{G_{;\kappa}G^{;\kappa}}{G}\nonumber\\
&&=-\frac{8\pi
G}{c^{2}}\bigg(T-\omega\frac{m^{2}k^{2}}{c^{2}}\phi_{\mu}\phi^{\mu}\bigg),\\
\label{m}
&&\frac{1}{m}m^{;\nu}_{~;\nu}-\frac{1}{m^{2}}m^{;\nu}m_{;\nu}
-\frac{1}{Gm}G^{;\nu}m_{;\nu}=-\frac{16\pi G}{c^{4}}\omega
m^{2}k^{2}\phi_{\mu}\phi^{\mu},\\
\label{O} &&\omega^{;\nu}_{~;\nu}-\frac{1}{G}G^{;\nu}\omega_{;\nu}
=\frac{16\pi
G}{c^{4}}\bigg(\frac{1}{4}B^{\mu\nu}B_{\mu\nu}+\frac{1}{2}m^{2}k^{2}\phi_{\mu}\phi^{\mu}\bigg).
\end{eqnarray}

\subsection[]{Perturbation of MSTVG}

Following the approach in \cite{b14,b50}, we deal with MSTVG in the
form of a Taylor expansion in the parameter $\epsilon\equiv1/c$,
similar to the expansion of the metric tensor in Eqs. (\ref{g00}),
(\ref{g0i}) and (\ref{gij}). For the expansions of vector field,
\begin{eqnarray}
\phi_{0}&=&\overset{(0)}{\varphi}_{0}+\epsilon^{2}\overset{(2)}{\varphi_{0}},\\
\phi_{i}&=&\epsilon\overset{(1)}{\varphi_{i}}.
\end{eqnarray}
and for the expansion of scalar field $G$
\begin{eqnarray}
G&=&G_{0}(1+\xi)=G_{0}\bigg(1+\epsilon^{2}\overset{(2)}{G}+\epsilon^{4}\overset{(4)}{G}\bigg),\\
\theta(G)&=&\theta_{0}+\theta_{1}\xi+\frac{1}{2}\theta_{2}\xi^{2}+\cdots,\\
\theta(G)&=&\theta_{0}+\epsilon^{2}\theta_{1}\overset{(2)}{G}+\epsilon^{4}\bigg(\frac{1}{2}\theta_{2}\overset{(2)}{G}^{2}+\theta_{1}
\overset{(4)}{G}\bigg),\\
\frac{\mathrm{d}\theta}{\mathrm{d}G}&=&\frac{1}{G_{0}}\bigg(\theta_{1}+\epsilon^{2}\theta_{2}\overset{(2)}{G}\bigg),
\end{eqnarray}
where $G_{0}$ is the background value of scalar field $G$. The
expansions of others scalar fields are
\begin{eqnarray}
m&=&m_{0}\bigg(1+\epsilon^{2}\overset{(2)}{m}+\epsilon^{4}\overset{(4)}{m}\bigg),\\
\omega&=&\omega_{0}\bigg(1+\epsilon^{2}\overset{(2)}{\omega}+\epsilon^{4}\overset{(4)}{\omega}\bigg),
\end{eqnarray}
where $m_{0}$ and $\omega_{0}$ are respectively the background
values for the scalar fields $m$ and $\omega$.

\subsection[]{Gauge condition}
We use the gauge condition imposed on the component of the metric
tensor proposed by Kopeikin $\&$ Vlasov \cite{b19} as follows:
\begin{equation}
\bigg(\frac{G_{0}}{G}\sqrt{-g}g^{\mu\nu}\bigg)_{,\nu}=0.
\end{equation}
Noted that the scalar field $G$ in MSTVG is in the inverse ratio to
the scalar field $\phi$ in \cite{b19}. To 1PN order, this gauge
gives
\begin{equation}
\label{gauge1}
\varepsilon^{2}\bigg(\frac{1}{2}H_{,i}-\frac{1}{2}N_{,i}-H_{ik,k}-\overset{(2)}{G}_{,i}\bigg)=0,
\end{equation}
and
\begin{equation}
\label{gauge2}
\varepsilon^{3}\bigg(L_{k,k}-\frac{1}{2}H_{,t}-\frac{1}{2}N_{,t}+\overset{(2)}{G}_{,t}\bigg)=0.
\end{equation}
Through covariant divergence of Eq. (\ref{vector2}), we derive a
useful formula
\begin{equation}
\label{gauge3}
\overset{(0)}{\varphi}_{0,t}-\overset{(1)}{\varphi}_{k,k}=\frac{1}{m^{2}_{0}k^{2}\omega_{0}}(J^{0}_{,t}+J^{k}_{,k})+\mathcal
{O}(2).
\end{equation}

\subsection{First order post-Newtonian approximation for MSTVG}

Based on fields equations of Eqs. (\ref{R}), (\ref{vector2}),
(\ref{G}), (\ref{m}) and (\ref{O}), we obtain the result of MSTVG in
1PN by using the gauge conditions (\ref{gauge1}), (\ref{gauge2}) and
Eq. (\ref{gauge3}) as follows.

The equation for $N$ and $\overset{(2)}{G}$ are
\begin{equation}
\label{N} \Box N=-8\pi \mathcal {G}\sigma,
\end{equation}
\begin{equation}
\Box \overset{(2)}{G}=(1-\gamma)4\pi\mathcal {G}\sigma,
\end{equation}
where $\Box$ is the D'Alembert operator in the Minkowski spacetime
and Newton's constant $\mathcal{G}$ is related to the constant
$G_{0}$ by
\begin{equation}
\label{g} \mathcal {G}=\frac{4+\theta_{0}}{3+\theta_{0}}G_{0}.
\end{equation}

Metric $H_{ij}$ is
\begin{equation}
\label{Hij} \Box H_{ij}=-8\gamma\pi \mathcal {G}\sigma\delta_{ij},
\end{equation}
where we can define $H_{ij}\equiv \delta_{ij}V$ and
\begin{equation}
\label{gamma1} \gamma\equiv\frac{\theta_{0}+2}{\theta_{0}+4}.
\end{equation}
Appendix A gives mathematical details of the derivation of this
important formula. A rigorous presentation in Appendix B identifies
the parameter $\gamma$ defined in Eq. (\ref{gamma1}) is just the PPN
parameter with the same symbol.

With $\theta_{0}=-1$, we obtain $\gamma=1/3$ which just reduces to
STVG. Then, Eq. (\ref{g}) becomes
\begin{equation}
\mathcal {G}=\frac{2}{1+\gamma}G_{0}.
\end{equation}
It is evident that $\mathcal{G}=G_{0}$ when $\gamma=1$. The above
shows that only the scalar field $G$ enters the metric $N$ and
$H_{ij}$ and we have given a strict proof about $\gamma\neq1$ in
general cases of STVG by Chandrasekhar's approach. (see Appendix A
and B)

Other metric components in 1PN are respectively
\begin{equation}
\label{LIII} \Box L_{i}=8(1+\gamma)\pi \mathcal {G}\sigma_{i},
\end{equation}
\begin{eqnarray}
\label{LL} \Box L&=&-4\pi
\mathcal {G}\bigg[(3\gamma-2\beta-1)N\sigma-2(1-\gamma)\sigma_{kk}\nonumber\\
&&+(3\gamma+1)\overset{(0)}{\varphi}_{0}J^{0}
+\frac{1}{2}(1+\gamma)\omega_{0}\Delta\overset{(0)}{\varphi}^{2}_{0}\nonumber\\
&&+2\gamma\omega_{0}\overset{(0)}{\varphi}_{0}\Delta\overset{(0)}{\varphi}_{0}\bigg]-\frac{1}{2}\beta\Delta
N^{2},
\end{eqnarray}
where $\Delta=\nabla^{2}$ and
\begin{equation}
\label{beta}
\beta\equiv1+\frac{\theta_{1}(1-\gamma)^{3}}{8(1+\gamma)}=1+\frac{\theta_{1}}{2(\theta_{0}+3)(\theta_{0}+4)^{2}}.
\end{equation}
Appendix B identifies $\beta$ in Eq. (\ref{beta}) as the
corresponding PPN parameter with the same symbol. In that Appendix
all the other PPN parameters except $\gamma$ and $\beta$ are proved
all zero for MSVTG. The ten PPN parameters for MSVTG are listed in
Table \ref{gammabeta}.
\begin{table}
\caption{\label{gammabeta} PPN parameters in MSTVG.}
\begin{ruledtabular}
\begin{tabular}{cccccccccc}
 $\gamma$ & $\beta$ & $\xi$ & $\alpha_{1}$ & $\alpha_{2}$ & $\alpha_{3}$ & $\zeta_{1}$ & $\zeta_{2}$ & $\zeta_{3}$& $\zeta_{4}$\\
\hline
$\frac{\theta_{0}+2}{\theta_{0}+4}$ & $1+\frac{\theta_{1}}{2(\theta_{0}+3)(\theta_{0}+4)^{2}}$ &0 & 0 & 0 & 0&0&0&0&0\\
\end{tabular}
\end{ruledtabular}
\end{table}

Equations for the vector and other scalar fields are
\begin{equation}
\label{vector}
\Box\overset{(0)}{\varphi}_{0}-m^{2}_{0}k^{2}\overset{(0)}{\varphi}_{0}=-\frac{J^{0}}{\omega_{0}},
\end{equation}
\begin{equation}
\Box \overset{(2)}{m}=0,
\end{equation}
\begin{equation}
\Box \overset{(2)}{\omega}=0.
\end{equation}
In Eq. (\ref{vector}), if we consider a vacuum case ($J^{\mu}=0$),
we find that the speed of vector field is not equal to the velocity
of light when we assume a plane wave solution which may cause
chromatic dispersion in vacuum.

\subsection{Equations of motion for MSTVG in 1PN}

Based on Ref. \cite{b25}, the equations of motion (EOM) derived from
$T^{\mu\nu}_{;\nu}=0$ is equivalent to $G^{\mu\nu}_{;\nu}=0$ if EEP
is satisfied. However, STVG violates EEP due to the vector field.
So, equations of motion in MSTVG have to be derived from the
covariant divergence of the Einstein tensor
$G^{\mu\nu}_{;\nu}\equiv0$. On the other hand, according to
\cite{b25}, we can infer that the equations of motion in MSTVG must
include the contribution of the vector field besides the matter and
metric. The Einstein tensor $G^{\mu\nu}$ in MSTVG is
\begin{eqnarray}
R^{\mu\nu}-\frac{1}{2}g^{\mu\nu}R&=&\frac{8\pi
G}{c^{2}}\bigg[T^{\mu\nu}+\frac{\omega}{c^{2}}\bigg(B^{\mu}_{~\kappa}B^{\nu\kappa}-\frac{1}{4}g^{\mu\nu}B_{\lambda\kappa}B^{\lambda\kappa}\bigg)\nonumber\\
&&+\omega
\frac{m^{2}k^{2}}{c^{2}}\phi^{\mu}\phi^{\nu}-\frac{m^{2}k^{2}}{c^{2}}\omega\frac{1}{2}g^{\mu\nu}\phi_{\alpha}\phi^{\alpha}\bigg]\nonumber\\
&&+\frac{1}{2}\bigg(\theta(G)\frac{G^{;\mu}G^{;\nu}}{G^{2}}+4\frac{G^{;\mu}G^{;\nu}}{G^{2}}-2\frac{G^{;\mu\nu}}{G}\nonumber\\
&&-\frac{m^{;\mu}m^{;\nu}}{m^{2}}+\omega^{;\mu}\omega^{;\nu}\bigg)\nonumber\\
&&-\frac{1}{4}g^{\mu\nu}\bigg(\theta(G)\frac{G_{;\kappa}G^{;\kappa}}{G^{2}}+8\frac{G_{;\kappa}G^{;\kappa}}{G^{2}}
-4\frac{G_{;\kappa}^{;\kappa}}{G}\nonumber\\
&&-\frac{m_{;\kappa}m^{;\kappa}}{m^{2}}+\omega_{;\kappa}\omega^{;\kappa}\bigg).
\end{eqnarray}
Then, the Bianchi identities $G^{i\nu}_{;\nu}\equiv0$ yields the
momentum equation
\begin{eqnarray}
\label{EOM} &&\sigma_{i,t}+\sigma_{ik,k}-\frac{1}{2}\sigma
N_{,i}+\overset{(0)}{\varphi}_{0,i}J^{0}\nonumber\\
&&+\epsilon^{2}\bigg(\frac{1}{2}\sigma VN_{,i}+\sigma
L_{i,t}-\frac{1}{2}\sigma L_{,i}\nonumber\\
&&+\frac{5}{2}\sigma_{i}V_{,t}-\frac{1}{2}\sigma_{i}N_{,t}+\sigma_{k}L_{i,k}-\sigma_{k}L_{k,i}\nonumber\\
&&+\frac{5}{2}\sigma_{ik}V_{,k}-\frac{1}{2}\sigma_{ik}N_{,k}+\frac{1}{2}\sigma_{kk}N_{,i}-\frac{1}{2}\sigma_{kk}V_{,i}\nonumber\\
&&-V\overset{(0)}{\varphi}_{0,i}J^{0}-\overset{(1)}{\varphi}_{i,t}J^{0}+\overset{(2)}{\varphi}_{0,i}J^{0}\nonumber\\
&&+\overset{(1)}{\varphi}_{k,i}J^{k}-\overset{(1)}{\varphi}_{i,k}J^{k}-\overset{(1)}{\varphi}_{i}(J^{0}_{,t}+J^{k}_{,k})\bigg)=0.
\end{eqnarray}
$G^{0\nu}_{;\nu}\equiv0$ yields the continuity equation
\begin{eqnarray}
&&\sigma_{,t}+\sigma_{k,k}\nonumber\\
&&+\epsilon^{2}\bigg(\frac{3}{2}V_{,t}\sigma-N_{,t}\sigma+\frac{3}{2}V_{,k}\sigma_{k}-N_{,k}\sigma_{k}-\sigma_{kk,t}+\overset{(0)}{\varphi}_{0,k}J^{k}\nonumber\\
&&+\overset{(0)}{\varphi}_{0}(J^{0}_{,t}+J^{k}_{,k})\bigg)=0.
\end{eqnarray}

\subsection{N-body pointlike for MSTVG in 1PN}

Considering an N-body system of nonspinning point masses for
simplicity, we follow the notation adopted by \cite{b20} and use
the matter stress energy tensor as follows
\begin{equation}
c^{2}T^{\mu\nu}(\mathbf{x},t)=\sum_{a}\mu_{a}(t)\upsilon_{a}^{\mu}\upsilon_{a}^{\nu}\delta(\mathbf{x}-\mathbf{y}_{a}(t)),
\end{equation}
where $\delta$ denotes the three-dimensional Dirac distribution, the
trajectory of the $a$th mass is represented by $\mathbf{y}_{a}(t)$,
the coordinate velocity of the $a$th body are
$\mathbf{v}_{a}=\mathrm{d}\mathbf{y}_{a}(t)/\mathrm{d}t$ and
$\upsilon_{a}^{\mu}\equiv(c,\mathbf{v}_{a})$, and $\mu_{a}$ denotes
an effective time-dependent mass of the $a$th body defined by
\begin{equation}
\mu_{a}(t)=\bigg(\frac{M_{a}}{\sqrt{gg_{\rho\sigma}\frac{\upsilon_{a}^{\rho}\upsilon_{a}^{\sigma}}{c^{2}}}}\bigg)_{a},
\end{equation}
where subscript $a$ denotes evaluation at the $a$th body and $M_{a}$
is the constant Schwarzschild mass. Another useful notation is
\begin{equation}
\tilde{\mu}_{a}(t)=\mu_{a}(t)\bigg[1+\frac{\upsilon_{a}^{2}}{c^{2}}\bigg],
\end{equation}
where $\upsilon_{a}^{2}=\mathbf{v}_{a}^{2}$. Both $\mu_{a}$ and
$\tilde{\mu}_{a}$ reduce to the Schwarzschild mass at Newtonian
order: $\mu_{a}=M_{a}+\mathcal {O}(\epsilon^{2})$ and
$\tilde{\mu}_{a}=M_{a}+\mathcal {O}(\epsilon^{2})$. Then the mass,
current, and stress densities in Eqs. (\ref{sigma}), (\ref{sigmai})
and (\ref{sigmaij}) for the N point masses read
\begin{eqnarray}
\sigma&=&\sum_{a}\tilde{\mu}_{a}\delta(\mathbf{x}-\mathbf{y}_{a}(t)),\\
\sigma_{i}&=&\sum_{a}\mu_{a}\upsilon_{a}^{i}\delta(\mathbf{x}-\mathbf{y}_{a}(t)),\\
\sigma_{ij}&=&\sum_{a}\mu_{a}\upsilon^{i}_{a}\upsilon^{j}_{a}\delta(\mathbf{x}-\mathbf{y}_{a}(t)).
\end{eqnarray}
We now assume a ``fifth force" matter current
\begin{equation}
\label{J}
c\mathcal{J}^{\mu}(\mathbf{x},t)=\sum_{a}Q_{a}\upsilon_{a}^{\mu}\delta(\mathbf{x}-\mathbf{y}_{a}(t)),
\end{equation}
where the ``fifth force charge" is
$Q_{a}\equiv\kappa\sqrt{G}\omega\mu_{a}(t)$ which coupled with
ordinary mass through coupling function $\omega$, where $\kappa$ is
a dimensionless constant. This assumption implies that this charge
is proportional to ordinary mass as in \cite{SET}. Substituting Eq.
(\ref{J}) into Eq. (\ref{J1}), we obtain
\begin{eqnarray}
J^{0}&=&\sum_{a}\kappa\sqrt{G}\omega\mu_{a}\delta(\mathbf{x}-\mathbf{y}_{a}(t)),\nonumber\\
J^{i}&=&\sum_{a}\kappa\sqrt{G}\omega\mu_{a}\upsilon_{a}^{i}\delta(\mathbf{x}-\mathbf{y}_{a}(t)).
\end{eqnarray}
Then, we obtain the following form in Newtonian approximation,
\begin{equation}
N=2\Delta^{-1}\{-4\pi\mathcal {G}\sigma\}=2\sum_{a}\frac{\mathcal
{G}M_{a}}{r_{a}}+\mathcal {O}(2),
\end{equation}
and
\begin{eqnarray}
\overset{(0)}{\varphi}_{0}&=&\frac{1}{4\pi\omega_{0}}\int\frac{e^{-m_{0}k|\mathbf{x}-\mathbf{z}|}}{|\mathbf{x}-\mathbf{z}|}
J^{0}(\mathbf{z},t)\mathrm{d}^{3}\mathbf{z}\nonumber\\
&=&\frac{\kappa}{4\pi}\sqrt{\frac{(1+\gamma)\mathcal
{G}}{2}}\sum_{a}\frac{M_{a}e^{-m_{0}kr_{a}}}{r_{a}}+\mathcal {O}(2).
\end{eqnarray}
Through integration of Eq. (\ref{EOM}), we have EOM in Newtonian
approximation as follows
\begin{eqnarray}
\label{a} \frac{\mathrm{d}v^{i}_{a}}{\mathrm{d}t}&=&-\sum_{b\neq
a}\frac{\mathcal
{G}M_{b}}{r_{ab}^{3}}r^{i}_{ab}\bigg\{1-\frac{\kappa^{2}\omega_{0}(1+\gamma)}{8\pi}e^{-m_{0}kr_{ab}}\bigg(1
+m_{0}kr_{ab}\bigg)\bigg\} +\mathcal{O}(2).
\end{eqnarray}
From Eq. (\ref{a}), the gravitational potential in Newtonian
approximation for MSTVG is that
\begin{equation}
\label{potential} U(r)=-\sum_{b\neq a}\frac{\mathcal
{G}M_{b}}{r_{ab}}\bigg(1-\frac{k^{2}\omega_{0}(1+\gamma)}{8\pi}e^{-m_{0}kr_{ab}}\bigg).
\end{equation}
On the other hand, Fischbach and Talmadge \cite{LLR} provided the
following potential
\begin{equation}
\label{U} U(r)=-\sum_{b\neq a}\frac{\mathcal
{G}M_{b}}{r_{ab}}\bigg(1+\alpha e^{-\frac{r_{ab}}{\lambda}}\bigg),
\end{equation}
which includes a usual Newtonian gravitational potential and a
Yukawa-type one. Compared between Eq. (\ref{potential}) and Eq.
(\ref{U}), we define $m_{0}k=1/\lambda$ and
$\alpha=-\frac{\kappa^{2}(1+\gamma)}{8\pi}\omega_{0}$. Then, Eq.
(\ref{a}) becomes
\begin{eqnarray}
\label{a2} \frac{\mathrm{d}v^{i}_{a}}{\mathrm{d}t}&=&-\sum_{b\neq
a}\frac{\mathcal {G}M_{b}}{r_{ab}^{3}}r^{i}_{ab}\bigg\{1+\alpha
e^{-\frac{r_{ab}}{\lambda}}\bigg(1
+\frac{r_{ab}}{\lambda}\bigg)\bigg\} +\mathcal{O}(2).
\end{eqnarray}
If MSTVG could explain galaxy rotation curves without exotic dark
matter, it must have $\alpha>0$ from Eq. (\ref{a2}).

The next step is to work out $N$ and $V$ easily in 1PN approximation
as
\begin{eqnarray}
N&=&2\sum_{a}\frac{\mathcal {G}M_{a}}{r_{a}}
+\epsilon^{2}\bigg\{\sum_{a}\frac{\mathcal
{G}M_{a}}{r_{a}}\bigg[4v^{2}_{a}-(n_{a}v_{a})^{2}\bigg]
+2(2-3\gamma)\sum_{a}\sum_{b\neq
a}\frac{\mathcal {G}^{2}M_{a}M_{b}}{r_{a}r_{ab}}\nonumber\\
&&+\sum_{a}\sum_{b\neq a}\frac{\mathcal
{G}^{2}M_{a}M_{b}}{r^{2}_{ab}}(n_{a}n_{ab})\bigg[1+\alpha
e^{-r_{ab}/\lambda}\bigg(1+\frac{r_{ab}}{\lambda}\bigg)\bigg]\bigg\}+\mathcal{O}(3),\\
V&=&2\gamma\sum_{a}\frac{\mathcal {G}M_{a}}{r_{a}}+\mathcal {O}(2),
\end{eqnarray}
by the relation of $\tilde{\mu}_{a}$ that
\begin{eqnarray}
\tilde{\mu}_{a}&=&M_{a}\bigg\{1+\epsilon^{2}\bigg[\bigg(N-\frac{3}{2}V\bigg)_{a}
+\frac{3}{2}v^{2}_{a}\bigg]+\mathcal {O}(4)\bigg\}\nonumber\\
&=&M_{a}\bigg\{1+\epsilon^{2}\bigg[(2-3\gamma)\sum_{b\neq
a}\frac{\mathcal
{G}M_{b}}{r_{ab}}+\frac{3}{2}v^{2}_{a}\bigg]+\mathcal {O}(4)\bigg\},
\end{eqnarray}
where $r_{a}=|\mathbf{x}-\mathbf{y}_{a}|$ and
$r_{ab}=|\mathbf{y}_{a}-\mathbf{y}_{b}|$. For $L_{i}$,
\begin{equation}
\label{Li} L_{i}=-2(1+\gamma)\sum_{a}\frac{\mathcal
{G}M_{a}}{r_{a}}v^{i}_{a}.
\end{equation}
Because $\alpha$ in the Newtonian order is very tiny (see Table
\ref{Yukawa}), we omit any coupling terms in the magnitude of
$\alpha\epsilon^2$. Thus, it yields
\begin{eqnarray}
\label{L} L&=&2(3\gamma-2\beta-1)\sum_{a}\sum_{b\neq
a}\frac{\mathcal
{G}^{2}M_{a}M_{b}}{r_{a}r_{ab}}-2(1-\gamma)\sum_{a}\frac{\mathcal
{G}M_{a}}{r_{a}}v^{2}_{a}\nonumber\\
&&-2\beta\sum_{a}\sum_{b}\frac{\mathcal
{G}^{2}M_{a}M_{b}}{r_{a}r_{b}}+\mathcal {O}(1,\alpha).
\end{eqnarray}
At last, by integration of Eq. (\ref{EOM}), we obtain EOM for MSTVG
in 1PN:
\begin{eqnarray}
\label{Npoint}
\frac{\mathrm{d}v^{i}_{a}}{\mathrm{d}t}&=&-\sum_{b\neq
a}\frac{\mathcal {G}M_{b}}{r_{ab}^{2}}n^{i}_{ab}-\alpha\sum_{b\neq
a}\frac{\mathcal
{G}M_{b}}{r^{2}_{ab}}n^{i}_{ab}\bigg(1+\frac{r_{ab}}{\lambda}\bigg)e^{-r_{ab}/\lambda}\nonumber\\
&&+\epsilon^{2}\bigg\{2(\gamma+\beta)\sum_{b\neq a}\frac{\mathcal
{G}^{2}M^{2}_{b}}{r^{3}_{ab}}n^{i}_{ab}+(2\gamma+2\beta+1)\sum_{b\neq a}\frac{\mathcal {G}^{2}M_{a}M_{b}}{r^{3}_{ab}}n^{i}_{ab}\nonumber\\
&&+\sum_{b\neq a}\frac{\mathcal
{G}M_{b}}{r^{2}_{ab}}\bigg[-\gamma v^{2}_{a}-(1+\gamma)v^{2}_{b}+\frac{3}{2}(n_{ab}v_{b})^{2}+2(1+\gamma)(v_{a}v_{b})\bigg]n^{i}_{ab}\nonumber\\
&&+\sum_{b\neq a}\frac{\mathcal
{G}M_{b}}{r^{2}_{ab}}\bigg[2(1+\gamma)(n_{ab}v_{a})-(2\gamma+1)(n_{ab}v_{b})\bigg]v^{i}_{ab}\nonumber\\
&&+\sum_{b\neq a}\sum_{c\neq a, b}\frac{\mathcal
{G}^{2}M_{b}M_{c}}{r^{2}_{ab}}\bigg[(2\beta-1)\frac{1}{r_{bc}}+2(\beta+\gamma)\frac{1}{r_{ac}}
-\frac{r_{ab}}{2r_{bc}^{2}}(n_{bc}n_{ab})\bigg]n^{i}_{ab}\nonumber\\
&& -\frac{1}{2}(4\gamma+3)\sum_{b\neq a}\sum_{c\neq
a,b}\frac{\mathcal
{G}^{2}M_{b}M_{c}}{r_{ab}r^{2}_{bc}}n^{i}_{bc}\bigg\}+\mathcal
{O}(\epsilon^{4},\epsilon^{2}\alpha),
\end{eqnarray}
where $n^{i}_{ab}=(y_{a}^{i}-y_{b}^{i})/r_{ab}$,
$\mathbf{v}_{ab}=\mathbf{v}_{a}-\mathbf{v}_{b}$, and scalar products
are denoted with parentheses, e. g.,
$(n_{ab}v_{ab})=\mathbf{n}_{ab}\cdot\mathbf{v}_{ab}$. Eq.
(\ref{Npoint}) will return to Einstein-Infeld-Hoffmann (EIH)
equations of motion in the PPN formalism \cite{b26,b48} when we
eliminate the contribution of the vector field in MSTVG.

\subsection{Secular periastron precession of binary pulsar for MSTVG in 1PN}

When we consider only two body ($M_{1}$ and $M_{2}$) in Eq.
(\ref{Npoint}), EOM of a point-mass binary for MSTVG in 1PN yields
\begin{equation}
\ddot{\mathbf{r}}=\mathbf{a}_{N}+\mathbf{a}_{1PN},
\end{equation}
where
\begin{eqnarray}
\mathbf{a}_{N}&=&-\frac{\mathcal {G}M}{r^{2}}\bigg[1+\alpha\bigg(1+\frac{r}{\lambda}\bigg)\exp{(-\frac{r}{\lambda})}\bigg]\mathbf{n},\\
\mathbf{a}_{1PN}&=&-\frac{\mathcal
{G}M}{c^{2}r^{2}}\bigg\{\bigg[\bigg(\gamma+3\nu\bigg)v^{2}
-2\bigg(\gamma+\beta+\nu\bigg)\frac{\mathcal
{G}M}{r}-\frac{3}{2}\nu\dot{r}^{2}\bigg]\mathbf{n}
-2\bigg(1+\gamma-\nu\bigg)\dot{r}\mathbf{v}\bigg\},
\end{eqnarray}
where $M=M_{1}+M_{2}$, $\dot{r}=(n_{12}v_{12})$,
$\mathbf{v}=\mathbf{v}_{1}-\mathbf{v}_{2}$, $\nu=M_{1}M_{2}/M^{2}$,
$\mathbf{r}=\mathbf{r}_{12}$, $\mathbf{n}=\mathbf{n}_{12}$ and
$\mathbf{a}=\mathrm{d} \mathbf{v}_{12}/\mathrm{d}t$. Compared with
\cite{b44}, additional term comes from the Yukawa force, namely,
the contribution of the vector field in MSTVG. By the aid of the
averaging method \cite{b21}, the secular periastron advance for a
binary pulsar in 1PN is
\begin{equation}
\label{omega} \frac{\mathrm{d\omega}}{\mathrm{d}t}=\frac{1}{2}
\frac{n\alpha
p^{2}}{\lambda^{2}}e^{-p/\lambda}+(2+2\gamma-\beta)\frac{\mathcal
{G}Mn}{p}\epsilon^{2},
\end{equation}
where $n^{2}a^{3}=\mathcal {G}M$, $p=a(1-e^{2})$, $a$ is the
semi-major axis and $e$ is the eccentricity of the binary. Here the
periastron shift of Yukawa force is
\begin{equation}
\label{123}
\frac{\mathrm{d\omega}}{\mathrm{d}t}_{Yukawa}=\frac{1}{2}
\frac{n\alpha p^{2}}{\lambda^{2}}e^{-p/\lambda},
\end{equation}
which is different from the result given by \cite{b39}. In this
paper, we keep the angular momentum in the process of using the
averaging method instead of approximate treatments such as
$\exp(-r/\lambda)\approx1-r/\lambda$ \cite{b39} to calculate the
$\dot{\omega}$.

\section[]{Summary of parameters for MSTVG}
\begin{table}
\caption{\label{ppn}Summary of the PPN parameters $\gamma$ and
$\beta$ in several gravity theoties.}
\begin{ruledtabular}
\begin{tabular}{cccccccccc}
 Parameter &  GR   & ST & MST & STVG  & MSTVG  & ${\AE}$ & TeVeS\\
\hline $\gamma$ &  1 & $\frac{\omega_{0}+1}{\omega_{0}+2}$ &
$1-\frac{2\alpha^{2}_{0}}{1+\alpha^{2}_{0}}$ &$\frac{1}{3}$ &
$\frac{\theta_{0}+2}{\theta_{0}+4}$ & 1 & 1\\
$\beta$ &  $1$  &
$1$$+\frac{\omega_1}{(2\omega_0+3)(2\omega_0+4)^2}$ &
$1+\frac{\beta_{0}\alpha^{2}_{0}}{2(1+\alpha^{2}_{0})^{2}}$ & $1$ &
$1+\frac{\theta_{1}}{2(\theta_{0}+3)(\theta_{0}+4)^{2}}$  & $1$ & 1\\
\hline
Violation of&   No & No & Yes & Yes & Yes  & No &No\\
EEP?&&&&&&&\\
\end{tabular}
\end{ruledtabular}
\end{table}

There are four parameters in MSTVG. They are two PPN parameters
($\gamma$, $\beta$) and two Yukawa parameters ($\alpha$, $\lambda$).
In this section, we mainly discuss these four parameters.

\subsection{Review and discussion on PPN parameters}

For a slow motion and weak field, PPN formalism introduces 10
parameters in a post-Newtonian metric to include various gravity
theories \cite{b27,b28}. PPN formalism can be traced back to
Eddington-Robertson-Schiff formalism \cite{b51,b52,b53}, which
introduce two PPN parameters $\gamma$ and $\beta$ by only
considering one point mass. At the same time, these two parameters
were endowed with some kind of meaning. For example, $\gamma$
denotes the level of space curvature and $\beta$ can be treated as
described the nonlinearity in the superposition law \cite{b26}.
Table \ref{ppn} lists the results of $\gamma$ and $\beta$ in: (1)
General relativity (GR); (2) Scalar-Tensor theory
\cite{b54,b55}(ST, where $\omega_{0}$ and $\omega_{1}$ are
coefficients of the coupling function of scalar field $\phi$ :
$\theta(\phi)=\omega_{0} +\omega_{1}\xi+\mathcal {O}(\xi^{2})$ if
$\phi=\phi_{0}(1+\xi)$); (3) Multi-Scalar-Tensor theory \cite{b4}
(MST, where $\alpha_{0}=\partial\ln
A(\varphi_{0})/\partial\varphi_{0}$,
$\beta_{0}=\partial\alpha(\varphi_{0})/\partial\varphi_{0}$); (4)
Scalar-Tensor-Vector theory \cite{b7} (STVG); (5) Modified
Scalar-Tensor-Vector theory in this paper (MSTVG); (6)
Einstein-aether theory \cite{b29} (${\AE}$); (7) The
tensor-vector-scalar theory provided by \cite{b30} (TeVeS, the
results $\gamma$ and $\beta$ given by \cite{b31}).

For different theories listed in Table \ref{ppn}, EEP is satisfied
except MST, STVG and MSTVG. PPN formalism only works in validity of
EEP so that comparisons with PPN parameters of all kinds of gravity
theories, especially MST, STVG and MSTVG, is only phenomenological.
We are going to argue that even though the PPN formalism is only
valid under EEP, $\gamma$ should not depart from 1 too much in STVG.
Generally, the deflection of light can be used to test the parameter
$\gamma$. Equations of motion for photons in 1PN contain only two
metric coefficients: $N$ and $V$. For MSTVG, equations of motion for
photons in 1PN still contain the contribution of vector field
besides the two metric coefficients. When considering only one point
mass, equations of motion for photons in 1PN for MSTVG is as follows
\begin{eqnarray}
\label{lightequation1}
\frac{\mathrm{d}^{2}x^{i}}{\mathrm{d}t^{2}}&=&-\frac{\mathcal
{G}M}{r^{3}}r^{i}-\alpha\frac{\mathcal
{G}M}{r^{3}}r^{i}\bigg(1+\frac{r}{\lambda}\bigg)e^{-r/\lambda}\nonumber\\
&&+\epsilon^{2}\bigg[-\gamma\frac{\mathcal
{G}M}{r^{3}}\bigg|\frac{\mathrm{d}\mathbf{x}}{\mathrm{d}t}\bigg|^{2}r^{i}+2(1+\gamma)\frac{\mathcal
{G}M}{r^{3}}\bigg(\mathbf{r}\cdot\frac{\mathrm{d}\mathbf{x}}{\mathrm{d}t}\bigg)\frac{\mathrm{d}x^{i}}{\mathrm{d}t}\bigg],\\
\label{lightequation2}
0&=&g_{\mu\nu}\frac{\mathrm{d}x^{\mu}}{\mathrm{d}t}\frac{\mathrm{d}x^{\nu}}{\mathrm{d}t}.
\end{eqnarray}
Let us assume the Newtonian order solution of these equation is
\begin{equation}
x^{i}_{N}\equiv c\hat{n}^{i}(t-t_{0}),
\end{equation}
where $|\hat{n}^{i}|=1$ and light in the Newtonian order travels in
a straight line at constant speed $c$. Furthermore, we assume the
first order solution of these equation is
\begin{equation}
\label{solution} x^{i}\equiv c\hat{n}^{i}(t-t_{0})+x^{i}_{p},
\end{equation}
where $\hat{n}^i$ is the initial emitting direction of a light
signal. Then, substituting Eq. (\ref{solution}) into Eqs.
(\ref{lightequation1})-(\ref{lightequation2}), we obtain
\begin{eqnarray}
\label{LS}
\frac{\mathrm{d}^{2}\mathbf{x}_{p}}{\mathrm{d}t^{2}}&=&-\frac{\mathcal
{G}M}{r^{3}}\bigg[1+\gamma+\alpha\bigg(1+\frac{r}{\lambda}\bigg)e^{-r/\lambda}\bigg]\mathbf{r}\nonumber\\
&&+2(1+\gamma)\frac{\mathcal
{G}M}{r^{3}}\mathbf{\hat{n}}(\mathbf{\hat{n}}\cdot\mathbf{r}),\\
\label{LS1}
\mathbf{\hat{n}}\cdot\frac{d\mathbf{x}_{p}}{\mathrm{d}t}&=&-\epsilon(1+\gamma)\frac{\mathcal
{G}M}{r}.
\end{eqnarray}
Followed the approach of Ref. \cite{b26}, by using Eqs.
(\ref{LS})-(\ref{LS1}), the light deflection angle up to 1PN
approximation for MSTVG is
\begin{equation}
\label{light} \Delta\phi=2\epsilon^{2}\frac{\mathcal
{G}M}{d}\bigg(1+\gamma+\alpha e^{-d/\lambda}\bigg),
\end{equation}
where $d$ represents the coordinate radius at the point of closest
approach of the ray. Then the deviation from GR is
\begin{equation}
\label{deltalight}
\delta_{\phi}\equiv\frac{\Delta\phi_{MSTVG}-\Delta\phi_{GR}}{\Delta\phi_{GR}}=\frac{1}{2}\bigg(\gamma-1+\alpha
e^{-d/\lambda}\bigg),
\end{equation}
which comes from two parts: PPN parameter $\gamma$ and the vector
field. If $\gamma=1/3$, the case of Eq. (\ref{light}) and
(\ref{deltalight}) will return to STVG \cite{b7}. Here, one would
claim that the part from the vector field contributes to remaining
$2/3$ which could make up $\gamma\neq1$ in STVG. However, this is
not the case. When considering the Lense-Thirring drag on the orbit
of an earth-satelite, the longitude of the ascending node $\Omega$
and the argument of pericenter $\omega$ vary with time by using the
averaging method \cite{b21}:
\begin{eqnarray}
\label{LT1}
\frac{\mathrm{d}\Omega}{\mathrm{d}t}&=&\frac{(1+\gamma)\mathcal{G}\epsilon^{2}J_{earth}}{a^{3}(1-e^{2})^{3/2}},\\
\label{LT2}
\frac{\mathrm{d}\omega}{\mathrm{d}t}&=&-\frac{2(1+\gamma)\mathcal{G}\epsilon^{2}J_{earth}}{a^{3}(1-e^{2})^{3/2}}\cos
i,
\end{eqnarray}
where $J_{earth}$ is the angular momentum of the earth and $i$ is
the inlination. Eqs. (\ref{LT1}) and (\ref{LT2}) return to GR when
$\gamma=1$. In Lense-Thirring effect, it worthy of note that
$\mathrm{d}\Omega/\mathrm{d}t$, namely Eq. (\ref{LT1}), only depends
on the force which is perpendicular to the orbital plane. In STVG,
however, the Yukawa force can not affect
$\mathrm{d}\Omega/\mathrm{d}t$ because it is a radial force in the
orbital plane. For this reason, $\mathrm{d}\Omega/\mathrm{d}t$ can
limit $\gamma$ very closely even though STVG violate the EEP. This
is the reason why we have to modified STVG.

Now, we mainly focus on the three gravity theories that include a
vector field. They are respectively ${\AE}$ \cite{b29}, TeVeS
\cite{b30} and STVG in this paper. ${\AE}$ is a tensor-vector
theory in which the vector field is massless, unitary and time like.
This theory investigates preferred frames and violation of Lorentz
invariance. There are five gravitational and aether wave modes for
${\AE}$-theory \cite{b63}. Recently, Xie and Huang \cite{b33}
presented a 2PN approximation of ${\AE}$-theory and found that the
linearized waves with the spin-0 and spin-1 modes in ${\AE}$ will
propagate with infinite velocities if the 2PN light deflection angle
in ${\AE}$-theory is identical with that of GR. MOdified Newtonian
Dynamics (MOND) \cite{b34,b35,b36} has gained recognition as a
successful scheme for explaining galaxy rotation curves without
invoking dark matter. TeVeS is a scenario of relativistic MOND,
which includes a unit massless timelike vector and a scalar field.
In TeVeS, modified Newtonian acceleration is from two positive
dimensionless parameters $K$ and $k$ but not scalar or vector field.
TeVeS passes the usual solar system tests and provides a specific
formalism for constructing cosmological models. For STVG, there are
three scalar fields, one vector field of rest mass $m$ besides the
metric field. Both STVG and TeVeS try to explain galaxy rotation
curves without dark matter, but the modified Newtonian acceleration
in STVG is from the vector field.

For MSTVG in 1PN, other eight PPN parameters except $\gamma$ and
$\beta$ are all zero. Measurement of the deflection of light passing
Jupiter by Very Long Baseline Interferometer (VLBI) gives
$\gamma-1=(-1.7\pm4.5)\times10^{-4}$ \cite{b42}. The most precise
measurement of $\gamma$ comes from the Cassini experiment by Doppler
tracking, which gives $\gamma-1=(2.1\pm2.3)\times10^{-5}$
\cite{b41}. By comparing the masses of 15 elliptical lensing
galaxies from Sloan Lens Advanced Camera for Surveys (SLACS) on
kiloparsec scales (about $10^{19}$m), Bolton et al. \cite{b32}
constrain $\gamma=0.98\pm0.07$ in $1\sigma$ confidence level (CL).
For $\beta$, although its accuracy level is lower than that of
$\gamma$, some experiments give the limits of $\beta$. For example,
$\beta-1=3\times10^{-3}$ comes from the perihelion shift of Mercury
\cite{b43} and $\beta-1=2.3\times10^{-4}$ comes from the Nordtvedt
effect \cite{b64,b65}. Tests of $\gamma$ and $\beta$ are basically
going on the solar system scale. However, according to present-day
experiment data, the values of these two parameters ($\gamma$ and
$\beta$) are independent of the scale of the testing systems. In
order to obtain $\gamma=1$ and $\beta=1$, the parameter $\theta_{0}$
in MSTVG must go to infinity with $\theta_{1}$ growing slower than
$\theta_{0}^{3}$.

\subsection{Review and discussion on Yukawa parameters}

\begin{table*}
 \caption{\label{Yukawa} Yukawa parameters $\alpha$ and
$\lambda$ in macroscopical range.}
\begin{ruledtabular}
\begin{tabular}{cccc}
 & $|\alpha|$ & $\lambda$ &Ref.\\
\hline
SDSS & $0.35\pm0.9$ & 10Mpc/h$<\lambda<$100Mpc/h &\cite{2DF}\\
2dFGRS & $0.025\pm1.7$ & 10Mpc/h$<\lambda<$100Mpc/h &\cite{2DF}\\
Pioneer anomaly 10/11, & & &\\
Galileo,and Ulysses data & $10^{-3}$ & $200$AU &\cite{PA}\\
Long range limit& $>10^{-4}$ & $>70$AU &\cite{LRL}\\
Pioneer anomaly (STVG) & $10^{-3}$ & $47\pm1$AU&\cite{CPA}\\
Constraint from Sun (STVG)& $10^{-8}$ & $10^{15}$m &\cite{SET}\\
Constraint from Earth (STVG)& $10^{-13}$ & $10^{13}$m & \cite{SET}\\
Planetary data(EPM2004)& $10^{-12}\sim10^{-13}$ & $1$AU &\cite{PD}\\
Constraint on solar system & $(0.3\pm2.7)\times10^{-11}$ &
$0.2\pm0.4$AU & \cite{b39}\\
planetary motions& $<10^{-9}$ & $<0.18$AU&\cite{solar}\\
LLR & $>10^{-10}$ & $10^{7}$m$<\lambda<10^{13}$m&\cite{LLR,LLR1}\\
planetary data& $>10^{-9}$ & $>10^{9}$m &\cite{LLR}\\
Constraint on intermediate-range gravity & $\approx10^{-8}$ &
$1.2\times10^{7}$m$<\lambda<3.8\times10^{8}$m & \cite{b40}\\
LAGEOS-lunar & $10^{-8}<|\alpha|<10^{-5}$ & $10^{5}$m$<\lambda<10^{9}$m &\cite{LLR}\\
Earth-LAGEOS & $10^{-6}<|\alpha|<10^{-3}$ & $10^{4}$m$<\lambda<10^{6}$m &\cite{LLR}\\
\end{tabular}
\end{ruledtabular}
\end{table*}

Since Fischbach et al. \cite{b38} suggest a possible gravity-like
``fifth" fundamental force in macroscopic scale, it evokes some
interest in many theories which intend to unify gravity with other
known forces. The presence of the fifth force could be detected by
searching for apparent deviations from Newtonian gravity. For
instance, the fifth force would arise from the exchange of a new
ultra-light boson which coupled to ordinary matter with a strength
comparable to gravity. Typically, through added a hypothetical
Yukawa force to the Newtonian potential, this modified potential per
mass takes the form:
\begin{equation}
\label{Y} U(r)=-\frac{\mathcal{G}M}{r}\bigg(1+\alpha
e^{-r/\lambda}\bigg),
\end{equation}
where $\alpha$ represents the strength of the Yukawa coupling, and
$\lambda$ represents its length scale.

In Table \ref{Yukawa}, we list some results about $\alpha$ and
$\lambda$ in macroscopical range from 100Mpc to $10^{4}$m. Fischbach
and Talmadge \cite{LLR} consider the model of Eq. (\ref{Y}) in
astronomical tests which provide tight constraints on Yukawa
parameters ($\alpha$ and $\lambda$). Typically, these results are
based on testing $G(r)M_{\odot}$ values deduced for different
planets through the following equation
\begin{equation}
\label{nnp} \mathbf{a}=-\frac{G(r)M}{r^{3}}\mathbf{r},
\end{equation}
where
\begin{equation}
\label{nnp1}
G(r)=\mathcal{G}\bigg[1+\alpha\bigg(1+\frac{r}{\lambda}\bigg)e^{-r/\lambda}\bigg].
\end{equation}
Planetary data gives $|\alpha|>10^{-9}$ and $\lambda>10^{9}$m (see
Table \ref{Yukawa}). The constraints at larger ranges from
laboratory, geophysical, and astronomical data are essentially
unchanged (for detail, see Figure 2.13 of \cite{LLR}). As to the
constraint from LLR, Adelberger et al.\cite{LLR1} updated the result
to include recent LLR data and gave $|\alpha|>10^{-10}$ and
$10^{7}<\lambda<10^{13}$m.

By using two large-scale structure surveys: the Sloan Digital Sky
Survey (SDSS) and the two-degree Field Galaxy Redshift Surveys
(2dFGRS) on megaparsec scales, Sealfon et al. \cite{2DF} considered
two models about Poission equation which deviated from gravitational
inverse-square law to constraint $\alpha$ with marginalized over
$\lambda$ from $10$Mpc/h to $100 $Mpc/h. Where h is the value of
Hubble constant in units of $100 $km/s/Mpc. The potential of the
first model is
\begin{equation}
\label{sdss}
\Phi(\mathbf{r})=-\mathcal{G}\int\mathrm{d}^{3}r'\frac{\rho(\mathbf{r}')}
{|\mathbf{r}-\mathbf{r}'|}\bigg[1+\alpha\bigg(1-e^{-|\mathbf{r}-\mathbf{r}'|/\lambda}\bigg)\bigg].
\end{equation}
After integration of Eq. (\ref{sdss}), it has the same result as the
Eq. (\ref{Y}). On this large-scale structure scales,
$\alpha=0.025\pm1.7$ for 2dFGRS and $\alpha=-0.35\pm0.9$ for SDSS at
$68\%$ confidence level (CL) through fitting the power spectrum
measurements from SDSS and 2dFGRS.

Li and Zhao \cite{b40} consider Eqs. (\ref{nnp})-(\ref{nnp1}) and
constrain the Yukawa parameters. Using the earth-satellite
measurement of earth gravity, the lunar orbiter measurement of lunar
gravity, and lunar laser ranging measurement (\cite{b40}), the
result from constraint on the two Yukawa parameters are
$\alpha=10^{-8}-5\times10^{-8}$ and
$\lambda=1.2\times10^{7}$m$-3.8\times10^{8}$m.

Moffat and Toth \cite{SET} express the acceleration law in STVG as
\begin{equation}
\label{1} \mathbf{a}=-\frac{G_{eff}M}{r^{3}}\mathbf{r},
\end{equation}
where
\begin{equation}
\label{2}
G_{eff}=\mathcal{G}\bigg[1+\alpha-\alpha\bigg(1+\frac{r}{\lambda}\bigg)e^{-r/\lambda}\bigg].
\end{equation}
With observations or experiments performed within the solar system
or in Earthbound laboratories, the authors estimate
$\lambda=5\times10^{15}$m, $|\alpha|=3\times10^{-8}$ for the Sun,
and $\lambda=8.7\times10^{12}$m, $|\alpha|=9\times10^{-14}$ for the
Earth.

By analysis of EPM2004 ephemerides, Iorio \cite{PD} constrains on
the strength and rang of a Yukawa-like fifth force through
considering potential Eq. (\ref{Y}) and gives
$|\alpha|=10^{-12}-10^{-13}$, $\lambda\approx1$AU. By using the same
potential Eq. (\ref{Y}), Iorio \cite{solar} constrains on the range
and the strength of a Yukawa-like fifth force with planetary
perihelia by the EPM2004 ephemerides: $\lambda<0.18$AU and
$|\alpha|<10^{-9}$. Through considering Eqs. (\ref{1})-(\ref{2}),
Iorio \cite{b39} obtains $\lambda=0.2\pm0.4$AU and
$|\alpha|=(0.3\pm2.7)\times10^{-11}$ by constraint on perihelion
shift of Mercury.

With Pioneer $10/11$ launched, their radiometric tracking data have
consistently indicated the presence of a small anomaly which is
$a_{p}=(8.74\pm1.33)\times10^{-10}$m/s$^{2}$, directed toward the
Sun \cite{b56}. This apparent anomalous constant acceleration has
been similarly shown in Galileo and Ulysses data \cite{b58}. In
order to explain this anomaly, some of recent efforts are looking
for new physics. Based on this idea, many modified Newton inverse
square laws were provided. John et al. \cite{PA} ruled out a lot of
potential causes and considered a gravitational potential with
introducing an additional Yukawa force but this modified potential
is a little different from Eq. (\ref{Y}) and has the following form
\begin{equation}
\label{ya} V(r)=-\frac{\mathcal{G}M}{(1+\alpha)r}\bigg(1+\alpha
e^{-r/\lambda}\bigg).
\end{equation}
Through identifying the last term of Eq. (\ref{ya}) as the Pioneer
$10/11$, Galileo and Ulysses acceleration, John et al. \cite{PA}
obtain $\alpha=-1\times10^{-3}$ for $\lambda=200$AU.

Reynaud and Jaekel \cite{LRL} also discussed the relation between
long range tests of the Newton law and the anomaly recorded on
Pioneer $10/11$ probes through considering the potential of Eq.
(\ref{Y}). With a power expansion of Eq. (\ref{Y}) in terms of
$r/\lambda$, Reynaud and Jaekel \cite{LRL} obtain a constant
anomalous acceleration on the range of distances probed by Pioneer
$10/11$ only if $\lambda>10$AU ($\alpha=-\lambda^{2}/\Lambda^{2}$
and $\Lambda=6300$AU). It yields $\alpha<-10^{-4}$ or
$|\alpha|>10^{-4}$. In STVG, Brownstein and Moffat \cite{CPA}
considered the anomaly recorded on Pioneer $10/11$ probes through
Eqs. (\ref{1})-(\ref{2}) and obtained
$\alpha=(1.00\pm0.02)\times10^{-3}$, $\lambda=47\pm1$AU. For these
results, it can be seen that a long-range Yukawa deviation from the
Newton potential can be treated as a constant acceleration.

From above analysis, the Yukawa parameters depend on the scale of
testing system. In galactic scale, Yukawa parameters would play a
big role to explain flat rotation curve \cite{b8}. However, in the
outer solar system, Yukawa parameters could provide to a certain
extent the Pioneer anomalous constant acceleration. Then, in the
inner solar system, the strength coupled with gravitation about the
Yukawa force, $\alpha$, is very small and the range, $\lambda$, is
large. These results lead to the fifth force negligible in inner
solar system. If future experiments confirm nonentity of dark matter
in the universe, it will provide a possible existence of a fifth
force. Until then, the fifth force must work in galactic scale even
cosmological scale. From this point of view, constraint on the
Yukawa parameters ($\alpha$ and $\lambda$) in MSTVG has somewhat
practical significance.

In view of these discussion about the four parameters in MSTVG: PPN
parameters ($\gamma$, $\beta$) and the Yukawa parameters ($\alpha$,
$\lambda$), the two PPN parameters are independence of the system
scale in the current experiments. However, the case for the two
Yukawa parameters are not. In the next section, the orbital data of
pulsar binaries are used to fit the Yukawa parameters, in which the
PPN parameters will be fixed as $\gamma=\beta=1$.

\section[]{Constraints on the Yukawa parameters by double neutron star binaries in MSTVG}
\begin{table*}
 \centering
 \begin{minipage}{160mm}
  \caption{\label{pusar} Parameters in binary pulsars.}
  \begin{tabular}{@{}llllll@{}}
  \hline
  Pulsars & $e$ & $M$ $(M_{\odot})$ & $P$ (day) & $<\dot{\omega}>$ $(^{\circ}\mathrm{yr}^{-1})$ & Ref.\\
\hline
PSR B1913+16 & 0.6171338(4) & 2.8281(2) & 0.322997462727(5) & 4.226595(5) & \cite{1913+16}\\
PSR B1534+12 & 0.2736767(1) & 2.679(2) &
0.420737299153(4) & 1.755805(3) & \cite{1534+12} \\
PSR J0737-3039 & 0.0877775(9) & 2.58708(16) & 0.10225156248(5)
& 16.89947(68) & \cite{0734-3039} \\
PSR B2127+11C & 0.681395(2) & 2.71279(13) & 0.33528204828(5) &
4.4644(1)
& \cite{2127+11} \\
\hline
\end{tabular}
\end{minipage}
\end{table*}

\begin{figure}
\includegraphics[width=85mm,angle=0]{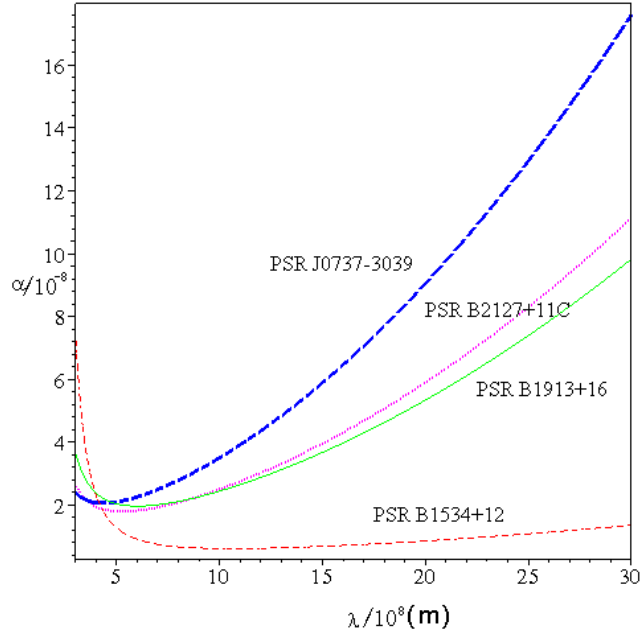}
 \caption{The plane of $\lambda$-$\alpha$ constrained by $\dot{\omega}$ of  four double neutron star
 binaries,
where the real line (green) denotes PSR B1913+16, the thinnest
dashed line (red) denotes PSR B1534+12,
 the thickest dashed line (blue) denotes PSR J0737-3039, and the dotted line (magenta) denotes PSR B2127+11C.
 It can be seen that the four curves are nearly cross at one point.}
  \label{fig1}
\end{figure}

\begin{figure}
\includegraphics[width=85mm,angle=0]{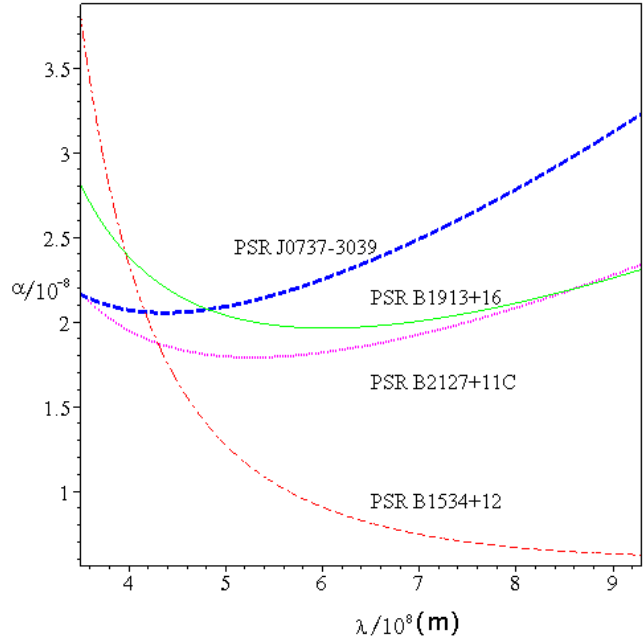}
 \caption{The enlarged diagram of the region around the cross point in Figure \ref{fig1}. The legend is the same as in Figure \ref{fig1}.
}
  \label{fig2}
\end{figure}

\begin{figure}
\includegraphics[width=85mm,angle=0]{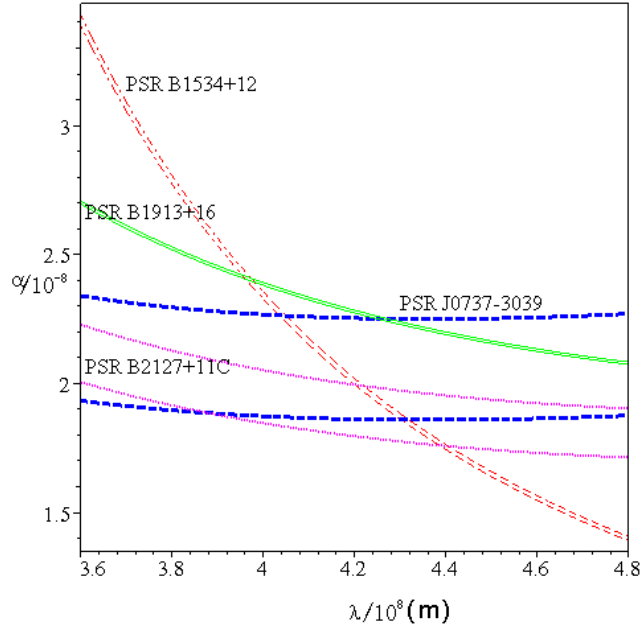}
 \caption{Each curve in Fig.\ref{fig2} is separated into two curves to show the bounds due to the observational $1\sigma$
  error. Please read the context for the details. Figure legend is the same as in Figure \ref{fig1}.
 }
  \label{fig3}
\end{figure}

\begin{figure}
\includegraphics[width=85mm,angle=0]{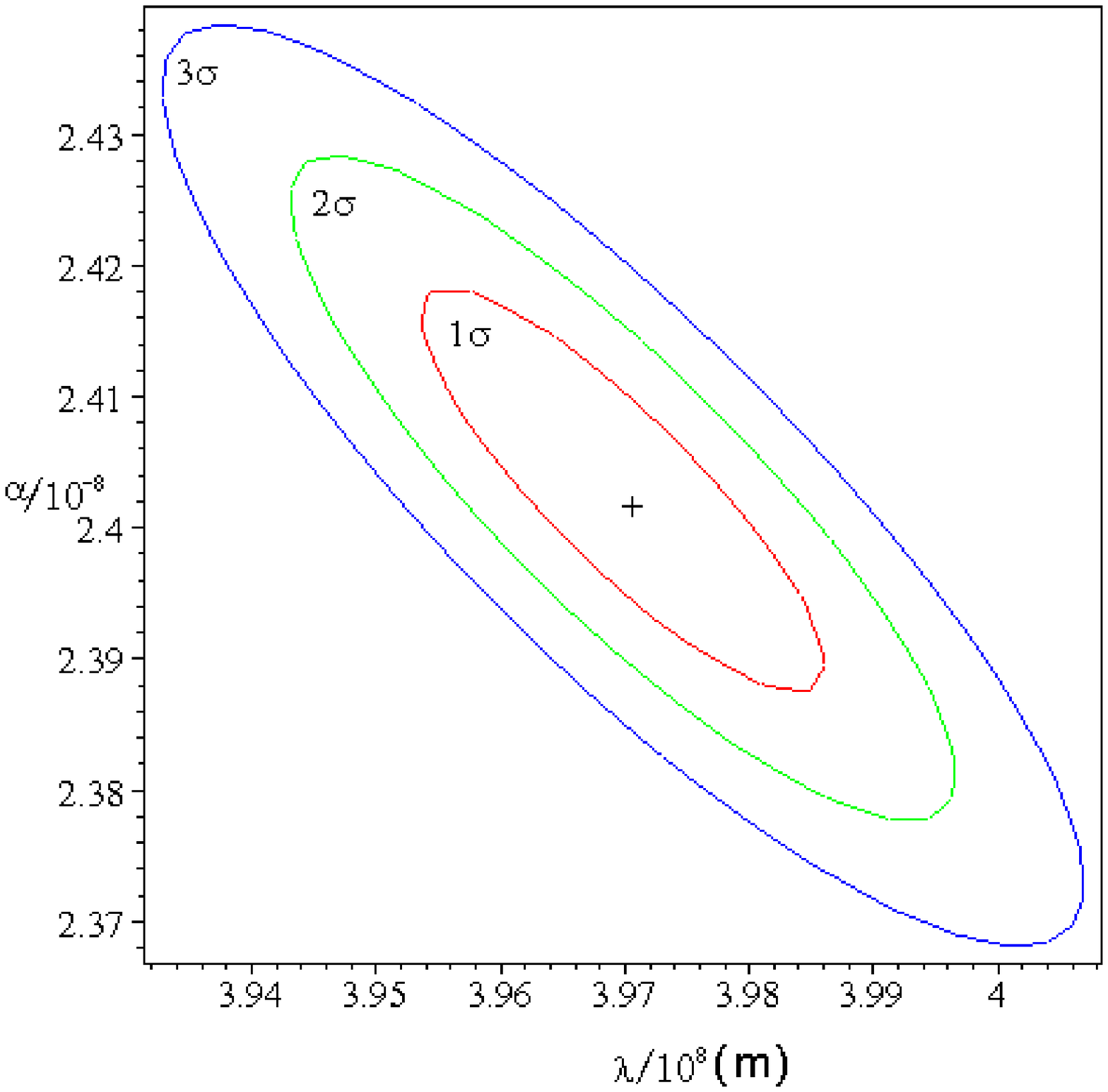}
 \caption{Confidence level contours of $68.3\%$ (red), $95.4\%$ (green) and $99.73\%$ (blue) in the plane of $\lambda$-$\alpha$
 constrained by four double neutron star binaries. The plus sign
 denotes the $68.3\%$ values for the Yukawa parameters of the MSTVG scenario which are:
$\lambda=(3.97\pm0.01)\times10^{8}$m
  and $\alpha=(2.40\pm0.02)\times10^{-8}$, the minimum value of $\chi^{2}$ is 21.47.}
  \label{fig4}
\end{figure}

Double neutron star binaries provide us more impressive tests of
general relativity than other systems. For example the fraction
which will merge due to gravitational wave emission is larger.
Besides, a neutron star is rather compact and its companion hardly
effect its shape. These systems are highly valuable for measuring
the effects of gravity and testing gravitational theories. Four
samples used in this paper are PSR B1913+16, PSR B1534+12, PSR
J0737-3039 and PSR B2127+11C which are listed in Table \ref{pusar}.
The numbers inside a pair of parentheses are the $1\sigma$ error of
its corresponding quantity.

PSR B1913+16 is discovered in 1974 by using the Arecibo 305m antenna
\cite{b24}. The orbit has evolved since the binary system was
initially discovered, in precise agreement with the loss of energy
due to gravitational wave emission predicted by Einstein's General
Theory of Relativity. PSR B2127+11C is located in the globular
cluster Messier 15. This system appears to be a clone of PSR
B1913+16 (see Table \ref{pusar}). For PSR B1534+12, its pulses are
significantly stronger and narrower than those of PSR B1913+16
\cite{1534+12}. PSR J0737-3039 was discovered during a pulsar
search carried out using a multibeam receiver with 64m radio
telescope in 2003 \cite{b59}. And we can see this system has
shorter orbital period, smaller eccentricity and larger periastron
advance. When we fix $\gamma=1$ and $\beta=1$, each $\dot{\omega}$
in Table \ref{pusar} will confine the values of $\alpha$ and
$\lambda$ in a curve through Eq.(\ref{omega}) (see Fig. \ref{fig1}
and Fig. \ref{fig2}).

In Fig. \ref{fig1}, the trajectories of four double neutron star
binaries in the plane of the Yukawa parameters have been shown. From
Fig. \ref{fig1}, we can see that the evolution of $\alpha-\lambda$
by PSR B1913+16 and PSR B2127+11C is very similar. For PSR B1534+12,
the evolution of $\alpha-\lambda$ is mostly flat. Besides, the
evolution of $\alpha-\lambda$ for PSR J0737-3039 is very steep. In
Fig. \ref{fig1}, we can also see that the four trajectories almost
meet together at $\lambda/10^{8}\simeq4$. Fig. \ref{fig2} is the
enlarged drawing of Fig. \ref{fig1} at a smaller scale. On the other
hand, Damour $\&$ Esposito-Far\`{e}se \cite{b4} constraint two 2PN
parameters of MST by using four different binary pulsars data in
$1\sigma$ confidence level. Based on the same method of \cite{b4},
we constrain the Yukawa parameters of MSTVG in $1\sigma$ confidence
level. Firstly, we plot an $1\sigma$ constraint imposed by double
neutron star binaries. Each set of binary data leads to a reduced
$\chi^{2}$:
$\chi^{2}_{binary}(\alpha,\lambda)=(\dot{\omega}_{theory}-\dot{\omega}_{observation})^{2}/\sigma^{2}_{observation}$,
equivalent to the $1\sigma$ constraint
$-\sigma_{observation}<\dot{\omega}_{theory}-\dot{\omega}_{observation}<\sigma_{observation}$.
The bounds by four double neutron star binaries allowed regions of
the $\lambda$-$\alpha$ plane are displayed in Fig. \ref{fig3}.
Clearly four binaries data favor only a small region of the Yukawa
parameters. To combine the constraints on $\alpha$ and $\lambda$
coming from different double neutron star binaries experiments, we
have added their individual $\chi^{2}$ as if they were part of a
total experiment with uncorrelated Gaussian errors like the analysis
method of \cite{b4}:
$\chi^{2}_{total}(\alpha,\lambda)=\chi^{2}_{1913+16}(\alpha,\lambda)+\chi^{2}_{1534+12}(\alpha,\lambda)
+\chi^{2}_{0737-3039}(\alpha,\lambda)+\chi^{2}_{2127+11C}(\alpha,\lambda)$.
Therefore, we plot the contour level
$\Delta\chi^{2}_{total}(\alpha,\lambda)=2.3$,
$\Delta\chi^{2}_{total}(\alpha,\lambda)=6.17$, and
$\Delta\chi^{2}_{total}(\alpha,\lambda)=11.8$, where
$\Delta\chi^{2}_{total}(\alpha,\lambda)=\chi^{2}_{total}(\alpha,\lambda)-(\chi^{2}_{total}(\alpha,\lambda))_{min}$,
defines respectively for two degrees of freedom the $68.3\%$,
$95.4\%$ and $99.73\%$ confidence levels represented in Fig.
\ref{fig4}. The $1\sigma$ fit values for the Yukawa parameters are
$\lambda=(3.97\pm0.01)\times10^{8}$m and
$\alpha=(2.40\pm0.02)\times10^{-8}$ with $\chi^{2}_{min}=21.47$.
With these results, we go back to discuss the light deflection for
Eq. (\ref{deltalight}). If we consider that a light ray which passes
the Sun at the solar radius, it yields
$\delta_{\phi}=3.33\times10^{-9}$. This means the deviation of MSTVG
from GR for defection of light may should be tested in the future
accuracy level.

For the same model of the Newtonian potential modified by the fifth
force (Eq. (\ref{Y})), we obtain the same results as in LLR
\cite{LLR,LLR1}, constraint on intermediate-range gravity
\cite{b40} and LAGEOS-lunar \cite{LLR} listed in Table
\ref{Yukawa}. It tells us that the limits of binary pulsar systems
on the Yukawa parameters for MSTVG are basically consistent with the
solar system. From the other view, it tells us that
$|2\gamma-\beta-1|$ almost equals to $0$. When considering three
parameters ($\alpha$, $\lambda$ and $|2\gamma-\beta-1|$) based on
Eq. (\ref{omega}) in $1\sigma$ level, we do not obtain any
reasonable results and need more precision binary pulsars data.

\section[]{Conclusions and prospects}

With relaxed the two assumptions taken in STVG, which actually hold
when the scalar field G is a constant field, it is shown that the
post-Newtonian parameter $\gamma\neq1$ by using Chandrasekhar's
approach and a modified scheme of scalar-tensor-vector gravity
theory (MSTVG) is then proposed by introducing a coupling function
of the scalar field $G$: $\theta(G)$. Started with the action in
MSTVG, the equations of motion of MSTVG in the first post-Newtonian
order (1PN) for general matter without specific equation of state
and N point masses are obtained. The secular periastron shift for
binary pulsar in 1PN is derived. From the results of MSTVG in 1PN,
there are four parameters: two PPN parameters $\gamma$ and $\beta$
and two Yukawa parameters $\alpha$ and $\lambda$. After pointed out
their independence of system scale for $\gamma$ and $\beta$,
discussion about the dependence of system scale for the Yukawa
parameters is touched. Furthermore, with the precondition of
$|2\gamma-\beta-1|=0$, $\alpha$ and $\lambda$ in MSTVG with applied
4 double neutron star binaries data (PSR B1913+16, PSR B1534+12, PSR
J0737-3039, PSR B2127+11C) are constrained:
$\lambda=(3.97\pm0.01)\times10^{8}$m and
$\alpha=(2.40\pm0.02)\times10^{-8}$. It has been shown that the
limits of binary pulsars systems on MSTVG are basically consistent
with the results from the solar system such as the earth-satellite
measurement of earth gravity, the lunar orbiter measurement of lunar
gravity, and lunar laser ranging measurement to constrain the fifth
force. For future applications in binary pulsars systems, it can
help us to distinguish between different gravity theories and MSTVG.
If MSTVG is proved to correct, besides success in solar system,
astrophysical and cosmological scales without dark matter, there
must exist a wave of vector field which does not equal to the light
speed. Besides, there also exist violations of the Einstein
equivalence principle in large scale in future experiments.

In Appendix B, we calculate the standard PPN parameters of the MSTVG, ignoring the vector field due to the coupling between it and the matter fields. When the vector field is included, some new super potentials might be introduced that would cause the appearance of new parameters and the numerical values of some PPN parameters might be affected. This is a subject of future research.

\begin{acknowledgments}
X.-M.Deng thanks the financial support in her research from the
Purple Mountain Observatory of China.
\end{acknowledgments}

\appendix

\section[Calculation of $\gamma$]{Calculation of $\gamma$}

In this appendix, we will give a detailed calculation of $\gamma$ in
MSTVG by using Chandrasekhar's approach. The contravariant
components of the metric tensor are
\begin{eqnarray}
\label{g00up}
g^{00}&=&-1-\epsilon^{2}N-\epsilon^{4}\bigg(N^{2}+L\bigg),\\
\label{g0iup}
g^{0i}&=&\epsilon^{3}L_{i},\\
\label{gijup} g^{ij}&=&\bigg(\delta_{ij}-\epsilon^{2}H_{ij}\bigg).
\end{eqnarray}
Accordingly,
\begin{equation}
    \sqrt{-g}=\bigg[1+\frac{1}{2}\epsilon^{2}(H-N)\bigg].
\end{equation}

We can now evaluate the Christoffel symbols with the aid of the
metric coefficients given in Eqs. (\ref{g00})-(\ref{gij}) and
(\ref{g00up})-(\ref{gijup}):
\begin{eqnarray}
    \Gamma^{0}_{00} & = & -\epsilon^{3}\frac{1}{2}N_{,t}-\epsilon^{5}\frac{1}{2}\bigg(NN_{,t}+L_{,t}+L_{k}N_{,k}\bigg),\\
    \Gamma^{0}_{0i} & = & -\epsilon^{2}\frac{1}{2}N_{,i}-\epsilon^{4}\frac{1}{2}\bigg(L_{,i}+NN_{,i}\bigg),\\
    \Gamma^{0}_{ij} & = & \epsilon^{3}\bigg(\frac{1}{2}H_{ij,t}-\frac{1}{2}L_{i,j}-\frac{1}{2}L_{j,i}\bigg),\\
    \Gamma^{i}_{00} & = & -\epsilon^{2}\frac{1}{2}N_{,i}+\epsilon^{4}\bigg(L_{i,t}-\frac{1}{2}L_{,i}+\frac{1}{2}N_{,k}H_{ik}\bigg),\\
    \Gamma^{i}_{0j} & = &\epsilon^{3}\bigg(\frac{1}{2}L_{i,j}-\frac{1}{2}L_{j,i}+\frac{1}{2}H_{ij,t}\bigg),\\
    \Gamma^{i}_{jk} & = &
    \epsilon^{2}\frac{1}{2}\bigg(H_{ij,k}+H_{ik,j}-H_{jk,i}\bigg).
\end{eqnarray}
Then, the general expression for the Ricci tensor is
\begin{eqnarray}
    \label{}
    R_{ij} & = &\epsilon^2\bigg( \frac{1}{2}N_{,ij}-\frac{1}{2}H_{ij,kk}+\frac{1}{2}H_{ik,kj}+\frac{1}{2}H_{jk,ki}-\frac{1}{2}H_{,ij}\bigg),\\
    R_{0i} & = &\epsilon^3\bigg(-\frac{1}{2}H_{,it}+\frac{1}{2}H_{ik,kt}-\frac{1}{2}L_{i,kk}+\frac{1}{2}L_{k,ki}\bigg),\\
    R_{00} & = & -\epsilon^2\frac{1}{2}N_{,kk}\nonumber\\
    & & +\epsilon^4\bigg(-\frac{1}{4}N_{,k}N_{,k}-\frac{1}{2}H_{,tt}
    +\frac{1}{2}N_{,lk}H_{lk}+\frac{1}{2}N_{,k}H_{kl,l}\nonumber\\
    &&\phantom{+\epsilon^4\bigg(}-\frac{1}{4}N_{,k}H_{,k}
    -\frac{1}{2}L_{,kk}+L_{k,kt}\bigg).
\end{eqnarray}

Turning to the components of the energy-momentum tensor, we find
\begin{eqnarray}
    T_{00}&=&\sigma-\epsilon^{2}(2N\sigma+\sigma_{kk}) ,\\
    T_{0i}&=&-\epsilon \sigma_{i},\\
    T_{ij}&=&\epsilon^{2}\sigma_{ij},
\end{eqnarray}
and
\begin{equation}
    T=-\sigma+\varepsilon^{2}\bigg(N\sigma+2\sigma_{kk}\bigg).
\end{equation}

The ($0,0$) component for $N$ from Eq. (\ref{R}):
\begin{equation}
\label{Nveri} \epsilon^{2}\bigg[-\frac{1}{2}\Box
N\bigg]=\epsilon^{2}\bigg[4\pi G_{0}\sigma+\frac{1}{2}\Box
\overset{(2)}{G}\bigg].
\end{equation}
The ($i,j$) component for $H_{ij}$ from Eq. (\ref{R}):
\begin{equation}
\epsilon^{2}\bigg[\frac{1}{2}N_{,ij}-\frac{1}{2}H_{ij,kk}+\frac{1}{2}H_{ik,kj}+\frac{1}{2}H_{jk,ki}-\frac{1}{2}H_{,ij}\bigg]
=\epsilon^{2}\bigg[4\pi
G_{0}\delta_{ij}\sigma-\overset{(2)}{G}_{,ij}-\frac{1}{2}\delta_{ij}\Box\overset{(2)}{G}\bigg].
\end{equation}
Using the gauge equation (\ref{gauge1}), we obtain
\begin{equation}
\label{Vveri} \epsilon^{2}\bigg[-\frac{1}{2}H_{ij,kk}\bigg]
=\epsilon^{2}\bigg[4\pi
G_{0}\delta_{ij}\sigma-\frac{1}{2}\delta_{ij}\Box\overset{(2)}{G}\bigg].
\end{equation}

For $\overset{(2)}{G}$ from Eq. (\ref{G})
\begin{equation}
\label{Gveri} \epsilon^{2}\bigg[(\theta_{0}+3)\Box
\overset{(2)}{G}\bigg]=\epsilon^{2}8\pi G_{0}\sigma
\end{equation}

Based on Eqs. (\ref{Nveri}) and (\ref{Gveri}), we obtain
\begin{equation}
\Box N=-\frac{\theta_{0}+4}{\theta_{0}+3}8\pi G_{0}.
\end{equation}
Based on Eqs. (\ref{Vveri}) and (\ref{Gveri}), we obtain
\begin{equation}
\Box H_{ij}=-\frac{\theta_{0}+2}{\theta_{0}+3}8\pi G_{0}\delta_{ij}.
\end{equation}
If we define $H_{ij}=\delta_{ij}V$, then we obtain
\begin{equation}
\Box V=-\frac{\theta_{0}+2}{\theta_{0}+3}8\pi G_{0},
\end{equation}
and
\begin{equation}
\label{gamma2} \gamma=\frac{V}{N}=\frac{\theta_{0}+2}{\theta_{0}+4}.
\end{equation}
When $\theta_{0}=-1$, it returns to the case of STVG and gives
$\gamma=1/3$. It is also explicit that $\gamma=1$ when the scalar
field G is a constant field. In this special case the terms with
$\overset{(2)}{G}$ in Eqs. (\ref{Nveri}) and (\ref{Vveri}) disappear
and $\gamma=1$ holds for both STVG and MSTVG. One might wonder at
Eq. (\ref{Gveri}), which does not allow $\overset{(2)}{G}$ to be
zero. But Eq. (\ref{Gveri}) comes from Eq. (\ref{G}), the field
equation for the scaler field G: Eq. (\ref{G}) does not exist when G
is not a variable field.

The next step is to identify the parameter $\gamma$ defined in Eq.
(\ref{gamma2}) as the corresponding PPN parameter. This is done in
Appendix B.

\section{Derivation of the PPN parameters for MSTVG}

In this appendix, we mainly derive the PPN parameters of MSTVG. For
this purpose, we must transfer our coordinate into the standard PPN
coordinate and then derive PPN parameters of MSTVG in comparison
with PPN formalism \cite{b26}. But, it is shown that MSTVG violates
EEP based on Eq. (\ref{action}) and can not be brought into the
standard PPN formalism. To compare with the PPN metric, however, we
provisionally ignore the vector field of MSTVG, which causes the
violation of EEP.

We consider the material composing the various bodies of the system
to behave like an ideal fluid. Following \cite{110}, the
energy-momentum tensor can be written in the following form in the
ideal fluid case
\begin{equation}
c^{2}T^{\mu\nu}=\rho(c^{2}+\Pi)u^{\mu}u^{\nu}+(g^{\mu\nu}+u^{\mu}u^{\nu})p,
\end{equation}
where $\rho$ denotes the rest-mass density, $p$ is the pressure,
$\Pi$ is the specific internal energy, and $u^{\mu}=dx^{\mu}/cd\tau$
is the dimensionless 4-velocity and we obtain in 1PN
\begin{eqnarray}
\label{T00}
T^{00}&=&\rho \bigg[1+\frac{1}{c^{2}}(\Pi+v^{2}+N)\bigg]+\mathcal{O}(c^{-4}),\\
\label{T0i}
T^{0i}&=&\rho \frac{v^{i}}{c}+\mathcal{O}(c^{-3}),\\
\label{Tij} T^{ij}&=&\frac{1}{c^{2}}(\rho
v^{i}v^{j}+p\delta_{ij})+\mathcal{O}(c^{-4}),
\end{eqnarray}
where $v^{i}$ is the coordinate velocity of the corresponding
material element.

From Eqs. (\ref{sigma})-(\ref{sigmai}) and (\ref{T00})-(\ref{Tij}),
we obtain
\begin{eqnarray}
\sigma&=&\rho\bigg[1+\frac{1}{c^{2}}(\Pi+2v^{2}+N)\bigg]+3\frac{p}{c^{2}}+\mathcal{O}(c^{-4}),\\
\sigma_{i}&=&\rho v^{i}+\mathcal{O}(c^{-2}),\\
\sigma_{kk}&=&\rho v^{2}+3p+\mathcal{O}(c^{-2}).
\end{eqnarray}
Then, we rewrite Eqs. (\ref{N}) and (\ref{LL}) with abandoning the
vector field
\begin{eqnarray}
\Delta\bigg[N+\epsilon^{2}L\bigg] &&\nonumber\\
=&&-8\pi\mathcal{G}\rho\nonumber\\
&&\epsilon^{2}\bigg\{-8\pi\mathcal{G}\rho\bigg[N+2v^{2}+\Pi+3\frac{p}{\rho}\bigg]\nonumber\\
&&-4\pi\mathcal{G}\rho\bigg[3\frac{\theta_{0}+2}{\theta_{0}+4}N-2\bigg(1+\frac{\theta_{1}}{2(\theta_{0}+3)(\theta_{0}+4)^{2}}\bigg)N-N\nonumber\\
&&-2\bigg(1-\frac{\theta_{0}+2}{\theta_{0}+4}\bigg)(v^{2}+3\frac{p}{\rho})\bigg]-\frac{1}{2}\bigg(1+\frac{\theta_{1}}{2(\theta_{0}+3)(\theta_{0}+4)^{2}}\bigg)\Delta
N^{2}\nonumber\\
&&+N_{,tt}\bigg\},
\end{eqnarray}
Then, we obtain
\begin{eqnarray}
\Delta N&=&-8\pi\mathcal{G}\rho,\\
\Delta
L&=&-4\pi\mathcal{G}\rho\bigg[3\frac{\theta_{0}+2}{\theta_{0}+4}N+N-2\bigg(1+\frac{\theta_{1}}{2(\theta_{0}+3)(\theta_{0}+4)^{2}}\bigg)N+\bigg(2\frac{\theta_{0}+2}{\theta_{0}+4}+2\bigg)v^{2}+2\Pi\nonumber\\
&&+6\frac{\theta_{0}+2}{\theta_{0}+4}\frac{p}{\rho}\bigg]
-\frac{1}{2}\bigg(1+\frac{\theta_{1}}{2(\theta_{0}+3)(\theta_{0}+4)^{2}}\bigg)\Delta
N^{2}+N_{,tt}.
\end{eqnarray}
and Eqs. (\ref{Hij}), (\ref{LIII}) are rewritten as
\begin{eqnarray}
\Delta H_{ij}&=&-8\frac{\theta_{0}+2}{\theta_{0}+4}\pi\mathcal{G}\rho\delta_{ij},\\
\Delta
L_{i}&=&8\bigg(1+\frac{\theta_{0}+2}{\theta_{0}+4}\bigg)\pi\mathcal{G}\rho
v^{i}.
\end{eqnarray}
After reference \cite{b26}, we define the following superpotentials
\begin{eqnarray}
U(\mathbf{x},t)&\equiv&\mathcal{G}\int\frac{\rho(\mathbf{x}',t)}{|\mathbf{x}-\mathbf{x}'|}\mathrm{d}^{3}x',\\
\chi(\mathbf{x},t)&\equiv&-\mathcal{G}\int\rho(\mathbf{x}',t)|\mathbf{x}-\mathbf{x}'|\mathrm{d}^{3}x',\\
\Phi_{1}(\mathbf{x},t)&\equiv&\int\frac{\rho(\mathbf{x}',t)v^{2'}}{|\mathbf{x}-\mathbf{x}'|}\mathrm{d}^{3}x',\\
\Phi_{2}(\mathbf{x},t)&\equiv&\int\frac{\rho(\mathbf{x}',t)U'}{|\mathbf{x}-\mathbf{x}'|}\mathrm{d}^{3}x',\\
\Phi_{3}(\mathbf{x},t)&\equiv&\int\frac{\rho(\mathbf{x}',t)\Pi'}{|\mathbf{x}-\mathbf{x}'|}\mathrm{d}^{3}x',\\
\Phi_{4}(\mathbf{x},t)&\equiv&\int\frac{p(\mathbf{x}',t)}{|\mathbf{x}-\mathbf{x}'|}\mathrm{d}^{3}x',\\
V_{i}(\mathbf{x},t)&\equiv&\int\frac{\rho(\mathbf{x}',t)v'_{i}}{|\mathbf{x}-\mathbf{x}'|}\mathrm{d}^{3}x',\\
W_{i}(\mathbf{x},t)&\equiv&\int\frac{\rho(\mathbf{x}',t)[\mathbf{v}'\cdot(\mathbf{x}-\mathbf{x}')](x^{i}-x^{i'})}{|\mathbf{x}-\mathbf{x}'|^{3}}\mathrm{d}^{3}x',\\
\Phi_{w}(\mathbf{x},t)&\equiv&\int\rho(\mathbf{x}',t)\rho(\mathbf{x}'',t)\frac{\mathbf{x}-\mathbf{x}'}{|\mathbf{x}-\mathbf{x}'|^{3}}
\bigg(\frac{\mathbf{x}'-\mathbf{x}''}{|\mathbf{x}-\mathbf{x}''|}-\frac{\mathbf{x}-\mathbf{x}''}{|\mathbf{x}'-\mathbf{x}''|}\bigg)
\mathrm{d}^{3}x'\mathrm{d}^{3}x'',\\
\mathfrak{A}(\mathbf{x},t)&\equiv&\int\frac{\rho(\mathbf{x}',t)[\mathbf{v}'\cdot(\mathbf{x}-\mathbf{x})]^{2}}{|\mathbf{x}-\mathbf{x}'|^{3}}\mathrm{d}^{3}x'.
\end{eqnarray}
From the above, we have the following relation
\begin{eqnarray}
\Delta\chi=-2U,\\
\chi_{,ti}=V_{i}-W_{i}.
\end{eqnarray}
With the above definition of the gravitational potentials, it yields
solution of the metric for MSTVG without the vector field in the
following forms
\begin{eqnarray}
\label{Nppn}
N&=&2U,\\
\label{Lppn}
L&=&-2\bigg(1+\frac{\theta_{1}}{2(\theta_{0}+3)(\theta_{0}+4)^{2}}\bigg)
U^{2}+(2\frac{\theta_{0}+2}{\theta_{0}+4}+2)\Phi_{1}\nonumber\\
&&+2\bigg[3\frac{\theta_{0}+2}{\theta_{0}+4}-2\bigg(1+\frac{\theta_{1}}{2(\theta_{0}+3)(\theta_{0}+4)^{2}}\bigg)+1\bigg]\Phi_{2}+2\Phi_{3}\nonumber\\
&&+6\frac{\theta_{0}+2}{\theta_{0}+4}\Phi_{4}-\chi_{,tt},\\
\label{Lippn}
L_{i}&=&-2\bigg(1+\frac{\theta_{0}+2}{\theta_{0}+4}\bigg)V_{i},\\
\label{vppn}
H_{ij}&=&2\frac{\theta_{0}+2}{\theta_{0}+4}\delta_{ij}U,
\end{eqnarray}

In order to obtain PPN parameters, we must transfer our coordinate
system into the standard PPN one. And then we could compare the
metric of MSTVG with the PPN metric in the standard PPN coordinate
system and finally derive the PPN parameters of MSTVG. When the
above is compared with the standard PPN metric (see Eqs.
(\ref{transferppn1})-(\ref{transferppn3}), the only superpotential
which does not appear in the PPN metric is $\chi_{,tt}$ in Eq.
(\ref{Lppn}). Therefore, it is necessary to transform the
coordinates to gauge off this term. Based on the Eq. (4.38) in Ref
\cite{b26}, an infinitesimal coordinate transformation is introduced
between the standard PPN coordinate system and ours:
\begin{equation}
\label{bianhuan}
x^{\mu}_{PN}=x^{\mu}+\epsilon^{2}\xi^{\mu}(x^{\alpha}),
\end{equation}
where
\begin{equation}
\label{Tr1}
\xi_{0}=\lambda_{1}\chi_{,0},~~~~\xi_{i}=\lambda_{2}\chi_{,i}.
\end{equation}
The relation between the metrics before and after the gauge
transformation, $g_{\mu\nu}$ and $\bar{g}_{\mu\nu}$ respectively,
are shown by  Eq. (4.46) in Ref \cite{b26}
\begin{eqnarray}
\label{Trgij}
\bar{g}_{ij}&=&g_{ij}-\epsilon^{2}\lambda_{2}\chi_{,ij},\\
\label{Trg0i}
\bar{g}_{0i}&=&g_{0i}-\epsilon^{3}(\lambda_{1}+\lambda_{2})\chi_{,ti},\\
\label{Trg00}
\bar{g}_{00}&=&g_{00}-\epsilon^{4}2\lambda_{1}\chi_{,tt}-\epsilon^{4}2\lambda_{2}(U^{2}+\Phi_{w}-\Phi_{2}),
\end{eqnarray}
Due to the spatial part of the PPN metric and our metric are all
diagonal and isotropic, we thus choose $\lambda_{2}=0$ through Eqs.
(\ref{Trgij}) and substitute Eq. (\ref{vppn}) into Eqs.
(\ref{Trgij}):
\begin{equation}
\label{transfer3}
\bar{g}_{ij}=\bigg(1+\epsilon^{2}2\frac{\theta_{0}+2}{\theta_{0}+4}
U\bigg)\delta_{ij}
\end{equation}
For Eqs. (\ref{Trg0i}) and (\ref{Trg00}), we have
\begin{eqnarray}
\label{transfer2}
\bar{g}_{0i}&=&-\epsilon^{3}2\bigg(1+\frac{\theta_{0}+2}{\theta_{0}+4}\bigg)V_{i}-\epsilon^{3}\lambda_{1}\chi_{,ti}\nonumber\\
&=&\epsilon^{3}\bigg[-\bigg(2\frac{\theta_{0}+2}{\theta_{0}+4}+2+\lambda_{1}\bigg)V_{i}+\lambda_{1}W_{i}\bigg],\\
\label{transfer1}
\bar{g}_{00}&=&g_{00}-\epsilon^{4}2\lambda_{1}\chi_{,tt}\nonumber\\
&=&-1+\epsilon^{2}2U+\epsilon^{4}\bigg[-2\bigg(1+\frac{\theta_{1}}{2(\theta_{0}+3)(\theta_{0}+4)^{2}}\bigg)
U^{2}+\bigg(2\frac{\theta_{0}+2}{\theta_{0}+4}+2\bigg)\Phi_{1}\nonumber\\
&&+2\bigg[3\frac{\theta_{0}+2}{\theta_{0}+4}-2\bigg(1+\frac{\theta_{1}}{2(\theta_{0}+3)(\theta_{0}+4)^{2}}\bigg)+1\bigg]\Phi_{2}+2\Phi_{3}+6\frac{\theta_{0}+2}{\theta_{0}+4}\Phi_{4}\nonumber\\
&&-(2\lambda_{1}+1)\chi_{,tt}\bigg],
\end{eqnarray}
by using Eqs. (\ref{Nppn}), (\ref{Lppn}) and (\ref{Lippn}). It is
noted that there is no existence of superpotential $\chi_{,tt}$ in
the standard PPN framework, we then obtain the following value
through Eq. (\ref{transfer1})
\begin{equation}
\lambda_{1}=-\frac{1}{2}
\end{equation}
When we substitute $\lambda_{2}=0$ and $\lambda_{1}=-\frac{1}{2}$
into Eq. (\ref{bianhuan}), the transformation between our reference
system and the standard PPN reference system is shown as
\begin{eqnarray}
\label{tpn}
t_{PN}&=&t+\epsilon^{4}\frac{1}{2}\chi_{,t}+\mathcal{O}(5),\\
\label{xpn} x^{i}_{PN}&=&x^{i}.
\end{eqnarray}
Through this transformation, our metric in the PPN coordinate system
become
\begin{eqnarray}
\label{MSTVG1}
\bar{g}_{ij}&=&\bigg(1+\epsilon^{2}2\frac{\theta_{0}+2}{\theta_{0}+4}
U\bigg)\delta_{ij},\\
\label{MSTVG2}
\bar{g}_{0i}&=&-\epsilon^{3}\frac{1}{2}\bigg(4\frac{\theta_{0}+2}{\theta_{0}+4}+3\bigg)V_{i}-\epsilon^{3}\frac{1}{2}W_{j},\\
\label{MSTVG3}
\bar{g}_{00}&=&-1+\epsilon^{2}2U+\epsilon^{4}\bigg\{-2\bigg(1+\frac{\theta_{1}}{2(\theta_{0}+3)
(\theta_{0}+4)^{2}}\bigg)U^{2}+\bigg(2\frac{\theta_{0}+2}{\theta_{0}+4}+2\bigg)\Phi_{1}\nonumber\\
&&+2\bigg[3\frac{\theta_{0}+2}{\theta_{0}+4}-2\bigg(1+\frac{\theta_{1}}{2(\theta_{0}+3)(\theta_{0}+4)^{2}}\bigg)+1\bigg]\Phi_{2}+2\Phi_{3}
+6\frac{\theta_{0}+2}{\theta_{0}+4}\Phi_{4}\bigg\},
\end{eqnarray}
by using Eqs. (\ref{transfer3})-(\ref{transfer1}).

On the other hand, with the above definition of the gravitational
potentials, the standard PPN metric \cite{b26} reads
\begin{eqnarray}
\label{transferppn1} \bar{g}_{ij}&=&(1+\epsilon^{2}2\gamma
U)\delta_{ij},\\
\label{transferppn2}
\bar{g}_{0i}&=&\epsilon^{3}\bigg[-\frac{1}{2}(4\gamma+3+\alpha_{1}-\alpha_{2}+\zeta_{1}-2\xi)V_{i}
-\frac{1}{2}(1+\alpha_{2}-\zeta_{1}+2\xi)W_{i}\bigg],\\
\label{transferppn3}
\bar{g}_{00}&=&-1+\epsilon^{2}2U+\epsilon^{4}\bigg\{-2\beta U^{2}-2\xi\Phi_{w}+(2\gamma+2+\alpha_{3}+\zeta_{1}-2\xi)\Phi_{1}\nonumber\\
&&+2(3\gamma-2\beta+1+\zeta_{2}+\xi)\Phi_{2}+2(1+\zeta_{3})\Phi_{3}+2(3\gamma+3\zeta_{4}-2\xi)\Phi_{4}\nonumber\\
&&-(\zeta_{1}-2\xi)\mathfrak{A}\bigg\}
\end{eqnarray}
Thus, the PPN parameters of MSTVG without the vector field have the
following forms by comparison between Eqs.
(\ref{MSTVG1})-(\ref{MSTVG3}) and
(\ref{transferppn1})-(\ref{transferppn3})
\begin{eqnarray}
\gamma&=&\frac{\theta_{0}+2}{\theta_{0}+4},\\
\beta&=&1+\frac{\theta_{1}}{2(\theta_{0}+3)(\theta_{0}+4)^{2}},\\
\xi&=&\alpha_{1}=\alpha_{2}=\alpha_{3}=\zeta_{1}=\zeta_{2}=\zeta_{3}=\zeta_{4}=0.
\end{eqnarray}

\newpage 
\bibliography{apssamp}

\end{document}